\documentclass[aps,prx,twocolumn,showpacs,superscriptaddress,amsmath,amssymb,amsfonts,floatfix,longbibliography]{revtex4-1}

\usepackage{graphicx}
\usepackage{dcolumn}
\usepackage{bm}
\usepackage{amssymb}
\usepackage[version=4]{mhchem}
\usepackage{microtype}
\newcolumntype{C}[1]{>{\centering\arraybackslash}p{#1}}
\usepackage{xfrac}
\usepackage{gensymb}
\usepackage{xcolor}
\usepackage{color}
\usepackage{multirow}
\usepackage{comment}
\usepackage{times}
\usepackage[varg]{txfonts}
\usepackage{mathrsfs}
\usepackage{amsmath,amssymb,bm,mathtools}
\usepackage{amsfonts}
\usepackage{booktabs}
\usepackage{natbib,hyperref}
\usepackage{scrextend}

\definecolor{cred}{RGB}{188,55,84}
\hypersetup{
	colorlinks=true, 
	linkcolor=blue, 
	citecolor=cred,
	urlcolor=cred,
}

\usepackage{soul}
\usepackage{epstopdf}
\usepackage{color}
\usepackage{array}
\usepackage{multirow}
\usepackage{tcolorbox}
\usepackage[percent]{overpic}
\usepackage{bbm}

\usepackage{widetable}
\usepackage{array}
\usepackage{xcolor,colortbl}
\usepackage{lipsum}

\newcommand{\ket}[1]{\ensuremath{|#1\rangle}}
\newcommand{\bra}[1]{\ensuremath{\langle#1|}} 
\newcommand{\avg}[1]{\langle #1 \rangle}

\newcommand{\vhat}[1]{\hat{#1}}

\newcommand{\tgo}{Tb$_2$Ge$_2$O$_7$}

\begin{document}

\title{Intertwined Magnetic Dipolar and Electric Quadrupolar Correlations in the Pyrochlore \texorpdfstring{Tb$_2$Ge$_2$O$_7$}{TEXT}}

\author{A.~M.~Hallas}
\affiliation{Department of Physics \& Astronomy and Quantum Matter Institute, University of British Columbia, Vancouver, British Columbia V6T 1Z1, Canada}
\email[Email: ]{alannah.hallas@ubc.ca}
\affiliation{Department of Physics and Astronomy, McMaster University, Hamilton, Ontario, L8S 4M1, Canada}
\affiliation{Canadian Institute for Advanced Research, MaRS Centre, West Tower 661 University Ave., Suite 505, Toronto, ON, M5G 1M1, Canada}

\author{W.~Jin}
\affiliation{Department of Physics and Astronomy, University of Waterloo, Waterloo, Ontario N2L 3G1, Canada}

\author{J.~Gaudet}
\affiliation{Department of Physics and Astronomy, McMaster University, Hamilton, Ontario, L8S 4M1, Canada}
\affiliation{Institute for Quantum Matter and Department of Physics and Astronomy, Johns Hopkins University, Baltimore, Maryland 21218, USA}
\affiliation{NIST Center for Neutron Research, National Institute of Standards and Technology, Gaithersburg, Maryland 20899, USA}

\author{E.~M.~Tonita}
\affiliation{Department of Physics and Astronomy, University of Waterloo, Waterloo, Ontario N2L 3G1, Canada}

\author{D. Pomaranski}
\affiliation{Department of Physics and Astronomy, University of Waterloo, Waterloo, Ontario N2L 3G1, Canada}

\author{C.~R.~C.~Buhariwalla}
\affiliation{Department of Physics and Astronomy, McMaster University, Hamilton, Ontario, L8S 4M1, Canada}

\author{M.~Tachibana}
\affiliation{National Institute for Materials Science, 1-1 Namiki, Tsukuba 305-0044, Ibaraki, Japan}

\author{N.~P.~Butch}
\affiliation{NIST Center for Neutron Research, National Institute of Standards and Technology, Gaithersburg, Maryland 20899, USA}

\author{S.~Calder}
\affiliation{Neutron Scattering Division, Oak Ridge National Laboratory, Oak Ridge, Tennessee 37831, USA}

\author{M.~B.~Stone}
\affiliation{Neutron Scattering Division, Oak Ridge National Laboratory, Oak Ridge, Tennessee 37831, USA}

\author{G.~M.~Luke}
\affiliation{Department of Physics and Astronomy, McMaster University, Hamilton, Ontario, L8S 4M1, Canada}

\author{C.~R.~Wiebe}
\affiliation{Department of Physics and Astronomy, McMaster University, Hamilton, Ontario, L8S 4M1, Canada}
\affiliation{Department of Chemistry, University of Winnipeg, Winnipeg, Manitoba, R3B 2E9 Canada}
\affiliation{Centre for Science at Extreme Conditions, University of Edinburgh, Edinburgh EH9 3FD, United Kingdom}

\author{J.~B. Kycia}
\affiliation{Department of Physics and Astronomy, University of Waterloo, Waterloo, Ontario N2L 3G1, Canada}

\author{M.~J.~P.~Gingras}
\affiliation{Department of Physics and Astronomy, University of Waterloo, Waterloo, Ontario N2L 3G1, Canada}

\author{B.~D.~Gaulin}
\affiliation{Department of Physics and Astronomy, McMaster University, Hamilton, Ontario, L8S 4M1, Canada}
\affiliation{Canadian Institute for Advanced Research, MaRS Centre, West Tower 661 University Ave., Suite 505, Toronto, ON, M5G 1M1, Canada}
\affiliation{Brockhouse Institute for Materials Research, McMaster University, Hamilton, Ontario, L8S 4M1, Canada}

\date{\today}


\begin{abstract}
We present a comprehensive experimental and theoretical study of the pyrochlore Tb$_2$Ge$_2$O$_7$, an exemplary realization of a material whose properties are dominated by competition between magnetic dipolar and electric quadrupolar correlations. The dipolar and quadrupolar correlations evolve over three distinct regimes that we characterize via heat capacity, elastic and inelastic neutron scattering. In the first regime, above $T^*=1.1$~K, significant quadrupolar correlations lead to an intense inelastic mode that cannot be accounted for within a scenario with solely magnetic dipole-dipole correlations. The onset of extended dipole correlations occurs in the intermediate regime, between $T^*=1.1$~K and $T_c = 0.25$~K, with the formation of a collective paramagnetic state characterized by extended ferromagnetic canted spin ice domains. Here, long-range order is impeded not only by the usual frustration operating in classical spin ice systems, but also by a competition between dipolar and quadrupolar correlations. Finally, in the lowest temperature regime, below $T_c=0.25$~K, there is an abrupt and significant increase in the dipole ordered moment. The majority of the ordered moment remains tied up in the ferromagnetic spin ice-like state, but an additional $\mathbf{k}=(0,0,1)$ antiferromagnetic order parameter also develops. Simultaneously, the spectral weight of the inelastic mode, which is a proxy for the quadrupolar correlations, is observed to drop, indicating that dipole order ultimately wins out. Tb$_2$Ge$_2$O$_7$ is therefore a remarkable platform to study intertwined dipolar and quadrupolar correlations in a magnetically frustrated system and provides important insights into the physics of the whole family of terbium pyrochlores.

\end{abstract}

\maketitle


\section{Introduction}

The concept of competing orders is a cornerstone of modern condensed matter physics. For example, the exotic properties and rich phase diagrams of cuprate, iron pnictide, organic, and heavy fermion superconductors originate from the competition between lattice, charge, and spin degrees of freedom and the compromised strongly correlated states that ensue. Highly frustrated magnetism is an exquisite setting to explore this sort of physics, with the aim of discovering novel forms of competing orders and exposing the mechanisms via which such competition is ultimately resolved. Even in insulating geometrically frustrated magnets, where only magnetic degrees of freedom are typically at play, much of the rich phenomenology arises from the competition between two or more magnetic orders~\cite{hallas2018experimental,rau2019frustrated,catuneanumagneticorders2015,catuneanupath2018,winterchallenges2016,wintermodels2017}.

Rare earth pyrochlores have proven to be a preferred setting to explore the physics of magnetic phase competition~\cite{gardner2010magnetic}. This is epitomized by several XY pyrochlores~\cite{ross2011quantum,robert2015spin,hallas2016universal,petit2017long,hallas2017phase,sarkis2019Yb2Ge2O7,schie2019Yb2Ti2O7}, which appear fine-tuned to lie near the cusp of adjacent magnetically ordered states, inducing proximate quantum spin liquid behavior~\cite{yan2017theory,hallas2018experimental,rau2019frustrated}. Most of this phase behavior has been successfully rationalized in the context of a nearest-neighbor Hamiltonian with anisotropic bilinear couplings between pseudospin $S~=~\sfrac{1}{2}$ degrees of freedom~\cite{ross2011quantum,savary2012order,wong2013phases,jaubert2015multiphase,yan2017theory}.
A crucially important ingredient for such a model is a crystal field ground state doublet that is well separated from the excited crystal field levels. In this work, we examine the intriguing case where this condition is no longer met, as realized in the terbium pyrochlores Tb$_2B_2$O$_7$ ($B =$~Ge, Ti, and Sn)
~\cite{rau2019frustrated,gingras2014quantum}. By lacking a clean separation of energy scales between the ion-ion interactions and the lowest energy excited crystal field level, the terbium pyrochlores belong to the broad and fascinating category of quantum many-body systems at ``intermediate coupling'', a famous example being organic conductors with triangular lattice geometries~\cite{kanoda2011Mott}.

There are two confounding factors that have inhibited 
understanding of the low temperature magnetism of the terbium pyrochlores. First and foremost is the aforementioned small energy separation of only 1.5~meV to the first excited crystal field level~\cite{gardner1999cooperative,gingras2000thermodynamic,gardner2010magnetic,rau2019frustrated}.
This is a significantly smaller value than in the other rare earth pyrochlores~\cite{bertin2012crystalfield}, which leads to a considerable admixing of this level with the ground state through the ion-ion interactions~\cite{kao2003understanding,molavian2007dynamically,molavian2009proposal,rau2019frustrated}. The quantitative simplicity and usefulness of a strictly bilinear anisotropic effective $S = \sfrac{1}{2}$ Hamiltonian in parameterizing the ion-ion interactions~\cite{lee2012generic,onoda2011quantumspinice} is therefore lost~\cite{wonglost2020}. Secondly, the crystal field ground state of the non-Kramers Tb$^{3+}$ ion is an $E_g$ doublet that is protected only by crystal symmetries~\cite{mirebeau2007magnetic,zhang2014neutron,princep2015crystal,ruminy2016crystal}. Consequently, the magnetic and lattice degrees of freedom can readily couple, giving rise to a large magneto-elastic
response~\cite{mamsurova1986low,ruff2007structural,ruff2010magnetoelastics,guitteny2013anisotropic,fennell2014magnetoelastic,ruminy2016sample,constable2017double,ruminy2019magnetoelastic}. The pseudospin $S = \sfrac{1}{2}$ associated with this non-Kramers ground state is highly anisotropic, with the local $z$ component behaving as a magnetic dipole while the transverse components behave as electric quadrupoles~\cite{lee2012generic,onoda2011quantumspinice}.

In the case of Tb$_2$Ti$_2$O$_7$, these complexities are further exacerbated by pronounced sensitivity to off-stoichiometry leading to dramatic sample dependence in its low temperature phase behavior~\cite{taniguchi2013long,kermarrec2015gapped}. While the earliest studies on this material revealed an absence of magnetic order down to 50 mK and hence a putative spin liquid state~\cite{gardner1999cooperative,gardner2003dynamic}, subsequent studies have suggested various forms of long- and short-range dipolar and quadrupolar ordered states~\cite{fritsch2013antiferromagnetic,taniguchi2013long,takatsu2016quadrupole,gritsenko2020changes}. Conversely, the low temperature phase behavior of Tb$_2$Sn$_2$O$_7$, which undergoes a long-range ordering transition into a canted spin ice state~\cite{mirebeau2005ordered}, appears at first sight rather simple. However, the apparently simple phase behavior of Tb$_2$Sn$_2$O$_7$ does not expose enough subtleties to allow the generic physics of the terbium pyrochlores to be understood. Our objective here is to reconcile these two extremes into one unified picture.


In this work, using specific heat measurements, elastic and inelastic neutron scattering, and theoretical modelling, we show how Tb$_2$Ge$_2$O$_7$ provides the crucial clues that have long been sought to unravel the physics of the terbium pyrochlores. We find that the magnetic correlations in Tb$_2$Ge$_2$O$_7$ evolve over three distinct temperature regimes, each heralded by a heat capacity anomaly, which we show to be a shared trait across the terbium pyrochlore family. However, Tb$_2$Ge$_2$O$_7$ has the cleanest and widest separation between these thermodynamic anomalies, enabling us to study its magnetic state with neutron scattering in each of the three temperature regimes. At temperatures above $T^* = 1.1$~K, we observe an intense inelastic mode that cannot be accounted for by correlations of the dipole moments of the Tb$^{3+}$ ions alone. We suggest that the brightness of this low energy mode arises from strongly correlated quadrupolar degree of freedom that are intertwined with the magnetic dipolar correlations. Passing through $T^* = 1.1$~K, there is a rapid onset of short-range ferromagnetic dipolar order, which is frustrated by the competing quadrupolar order. Finally, below $T_c=0.25$~K, Tb$_2$Ge$_2$O$_7$ undergoes a first order transition into a magnetically ordered state with both ferromagnetic and antiferromagnetic components. This indicates that dipolar order ultimately trumps quadrupolar order in this material. While a minimal $S = \sfrac{1}{2}$ model can describe the competition between magnetic dipolar phases and quadrupolar ones~\cite{onoda2011quantumspinice,lee2012generic}, it must be put aside in the present case as it leaves out how these competing orders can intertwine to produce complex phases. To address this challenge, we use mean-field theory and the random phase approximation to compute the phase diagram and the inelastic neutron scattering for a minimal model that incorporates coupled magnetic and quadrupolar degrees of freedom as well as the low-lying crystal field excitation. This allows us to identify a scenario and a set of possible competing phases that can underlie the phenomenology observed across the family of terbium pyrochlores.


\section{Sample Preparation \& Experimental Methods}

The difference in ionic radii between Tb$^{3+}$ and Ge$^{4+}$ is larger than can typically be accommodated by the pyrochlore lattice~\cite{subramanian1983oxide}. Thus, incorporating them into the cubic pyrochlore structure rather than the tetragonal pyrogermanate structure necessitates a high pressure synthesis method~\cite{shannon1968synthesis}. Stoichiometric quantities of Tb$_2$O$_3$ and GeO$_2$ were thoroughly mixed and then reacted in gold capsules using a belt type pressure apparatus at 6~GPa and 1300\degree C. After reacting for one hour, the samples were rapidly quenched to room temperature prior to the pressure being released. The resultant product is Tb$_2$Ge$_2$O$_7$ in the cubic pyrochlore phase (space group $Fd\bar{3}m$). Each $0.4-0.5$~g batch was individually x-rayed to confirm phase purity. The total sample mass was 3.9~g.

Heat capacity measurements were performed in a magnetically shielded dilution refrigerator in zero field ($< 0.001$ G). A 230 mg portion of the Tb$_2$Ge$_2$O$_7$ sample was pressed with 200 mg of high purity (99.9\%) silver powder, to improve thermal equilibration time. The pressed pellet was suspended in vacuum on four 6 $\mu$m diameter, 1 cm long nylon threads. A 10 k$\Omega$ metal film resistor and a 1 k$\Omega$ RuO$_2$ chip resistor~\cite{meisel1989thick} were mounted on the sample and used as a heater and thermometer, respectively. The heat capacity was determined using the thermal relaxation method~\cite{bachmann1972heat}. In this method, the sample was first heated to a temperature of 1 K, and the applied power was then removed, allowing the sample to cool back to the temperature of the dilution refrigerator cold stage. The specific heat was then determined from the slope of the cooling rate of the sample, which cooled for a period of 4 hours over the temperature span of the specific heat data.

Powder neutron diffraction measurements on Tb$_2$Ge$_2$O$_7$ were performed at the HB-2A beam line at the High Flux Isotope Reactor at Oak Ridge National Laboratory. A base temperature of 0.3~K was achieved using an orange cryostat with a $^3$He insert. Measurements were performed with a wavelength of 2.41~\AA, covering a wave vector, $Q$, range from 0.5 to 5.0 \AA$^{-1}$. The magnetic diffraction pattern at $0.3$~K was isolated by subtracting off a data set of equivalent statistics collected at $2$~K. The magnetic symmetry analysis was performed with SARAh~\cite{wills2000new} and Rietveld refinements were carried out using FullProf~\cite{rodriguez1993recent}.

Inelastic neutron scattering experiments on Tb$_2$Ge$_2$O$_7$ were performed on both the Disc Chopper Spectrometer (DCS) at the National Institute for Standards and Technology and SEQUOIA at the Spallation Neutron Source~\cite{granroth2010sequoia}. SEQUOIA is optimized for larger energy transfers, suitable for measuring the crystal electric field excitations, which typically span up to 100~meV in rare earth pyrochlores~\cite{gardner2010magnetic,bertin2012crystalfield}. Our measurements on SEQUOIA were carried out with an orange cryostat, over a temperature range of 2 K to 200 K, and with incident neutron energies of 8~meV, 30~meV, and 150~meV. The DCS operates at lower energy transfers, ideal for probing the collective magnetic excitations of a rare earth magnet. Our DCS measurements on Tb$_2$Ge$_2$O$_7$ were collected with incident energies of 3.3 meV and 1.3 meV, with an ICE dilution fridge giving a base temperature of 0.06~K. For the DCS experiment, our powder sample of Tb$_2$Ge$_2$O$_7$ was wrapped in Cu foil and loaded in 10 atm of He gas in a Cu sample can, to improve thermal equilibration at low temperatures. All inelastic data was reduced and analysed using the DAVE software suite~\cite{azuah2009dave}.


\section{Experimental Results}
\label{sec:experiment}


\subsection{Heat Capacity}

\begin{figure}[tbp]
\linespread{1}
\par
\includegraphics[width=3.2in]{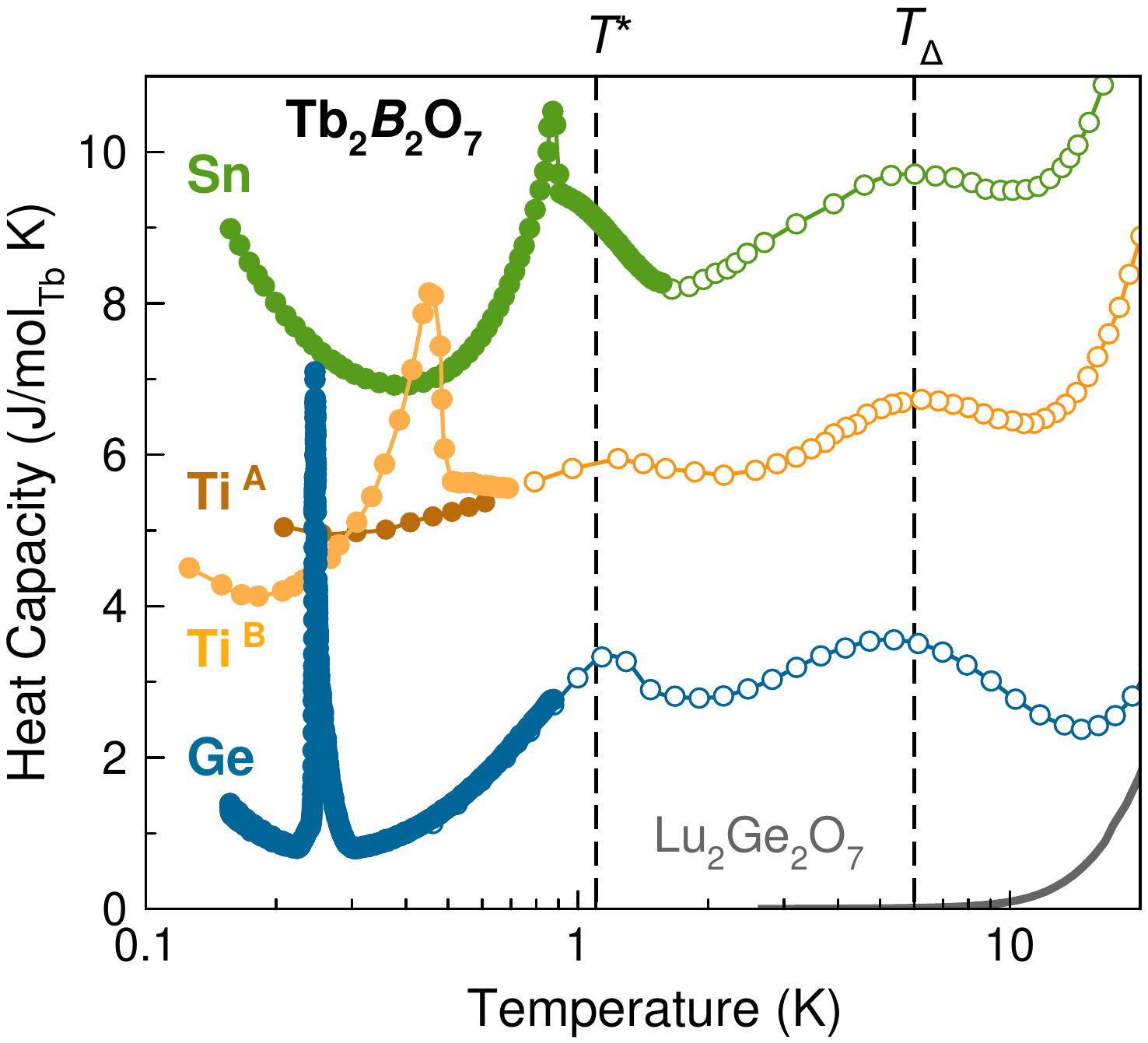}
\par
\caption{Heat capacity measurements for Tb$_2B_2$O$_7$ ($B=$~Ge, Ti, and Sn) on a logarithmic temperature scale covering more than two orders of magnitude. The higher temperature data sets given by the open symbols are reproduced from Refs.~\cite{hallas2014incipient} (Ge), \cite{gingras2000thermodynamic} (Ti), and \cite{chapuis2010evidence} (Sn). The lower temperature data sets given by the filled symbols are original to this work in the case of $B=$~Ge and reproduced from Refs.~\cite{kermarrec2015gapped} (Ti) and \cite{mirebeau2005ordered} (Sn). The data for $B =$~Ti and Sn have been offset vertically for clarity by 3 and 6~J/mol$_{\text{Tb}}$-K, respectively. Two representative low temperature data sets are included for $B=$~Ti, which exhibits profound sample dependence. All three members of the Tb$_2B_2$O$_7$ family exhibit a similar sequence of heat capacity anomalies. The two higher temperature of these, at $T_{\Delta} = 6$~K and $T^{*} = 1.1$~K, are relatively constant while the sharper, low temperature anomaly varies considerably, ranging from $T_{c} = 0.25$~K (Ge) to $T_{c} = 0.45$~K or altogether absent (Ti) to $T_{c}=0.87$~K (Sn). The solid gray curve is the heat capacity for non-magnetic Lu$_2$Ge$_2$O$_7$ from Ref.~\cite{li2016long}.}
\label{Tb_Cp} 
\end{figure}

To begin exploring the magnetic phase behavior of Tb$_2$Ge$_2$O$_7$, we first consider the temperature dependence of its magnetic specific heat, and compare it with that of Tb$_2$Sn$_2$O$_7$ and Tb$_2$Ti$_2$O$_7$. Heat capacity measurements for the three terbium pyrochlores, Tb$_2B_2$O$_7$ with $B=$~Ge, Ti, and Sn, are shown in Fig.~\ref{Tb_Cp}. Note that the temperature axis is on a logarithmic scale, and the data sets have been vertically offset for clarity. These data sets span more than two orders of magnitude in temperature and thus, different experimental setups are required for the high temperature (open symbols -- $^4$He or $^3$He cryostat) and low temperature (closed symbol -- dilution refrigerator) limits. Five of these data sets are taken from the published literature for Tb$_2$Sn$_2$O$_7$~\cite{chapuis2010evidence,mirebeau2005ordered}, Tb$_2$Ti$_2$O$_7$~\cite{gingras2000thermodynamic,kermarrec2015gapped}, and Tb$_2$Ge$_2$O$_7$~\cite{hallas2014incipient}. The low temperature thermodynamic properties of Tb$_2$Ti$_2$O$_7$ are exceedingly sensitive to low levels of defects. We therefore present two representative low temperature data sets for Tb$_{2+x}$Ti$_{2-x}$O$_{7+y}$ from Ref.~\cite{kermarrec2015gapped}: one that does not show an ordering transition below 1~K (sample A, $x = -0.0010(2)$) and one that does (sample B, $x = 0.0042(2)$). 
The low temperature heat capacity data for Tb$_2$Ge$_2$O$_7$ (filled blue circles) is original to the present work. This new data reveals a sharp and apparently first-order phase transition at $T_c=0.25$~K.

Our comparison of the heat capacity data for the terbium pyrochlores, Tb$_2B_2$O$_7$ with $B=$~Ge, Ti, and Sn, over more than two decades in temperature makes it clear that all three share a very similar set of thermodynamic anomalies. First, we observe that they all have a broad anomaly at $T_{\Delta} = 6$~K, which is a Schottky anomaly associated with the thermal population of the first excited crystal field level at approximately $1.5$~meV~\cite{gingras2000thermodynamic}. Next, we see that they each have a broad anomaly at approximately $T^* =1.1$~K. Then finally, all three exhibit a sharper heat capacity anomaly at low temperatures, and it is only this lowest temperature anomaly at $T_c$ which varies appreciably across the series. This lowest temperature anomaly is sharpest and at the lowest temperature for Tb$_2$Ge$_2$O$_7$ where $T_c = 0.25$~K. In the case of Tb$_2$Sn$_2$O$_7$, the anomalies associated with $T^*$ and $T_c$ are almost on top of each other -- nevertheless it is clear that a broad anomaly at $T^*$, appearing as a shoulder in this case, precedes the sharp anomaly at just lower temperature, $T_c = 0.87$~K. The three thermodynamic anomalies are thus widely separated for Tb$_2$Ge$_2$O$_7$ and poorly separated for Tb$_2$Sn$_2$O$_7$, with Tb$_2$Ti$_2$O$_7$ intermediate, and notwithstanding the aforementioned sensitivity of its existence on sample stoichiometry~\cite{taniguchi2013long}. The specific heat results strongly suggest that there is a shared character to the low temperature phase behavior across this family, in spite of any material dependent details.

Using the scaled heat capacity of the non-magnetic analog, Lu$_2$Ge$_2$O$_7$ (grey line in Fig.~\ref{Tb_Cp}), we can isolate the magnetic component of the specific heat for Tb$_2$Ge$_2$O$_7$. The calculated magnetic entropy release reaches $R\ln{(2)}$ by 3~K and $R\ln{(4)}$ by 20~K, as would be expected for a ground state doublet and a low-lying excited crystal field doublet at an energy $\sim 1.5$ meV (\emph{i.e.} 17.4 K). The entropy release associated with the transition at $T_c = 0.25$~K is approximately 10\% of $R\ln{(2)}$.


\subsection{Crystal Field Analysis and Single Ion Anisotropy}
\label{subsec:cf-analysis}

As is generally the case for rare earth magnets, single ion properties are the foundation upon which all other measures of the magnetic correlations are built~\cite{rau2019frustrated}. We therefore begin by determining the single ion crystal field energy spectrum in Tb$_2$Ge$_2$O$_7$, as was previously done for the other terbium pyrochlores~\cite{zhang2014neutron,princep2015crystal,ruminy2016crystal,mirebeau2007magnetic,zhang2014neutron}.

\begin{figure}[tbp]
\linespread{1}
\par
\includegraphics[width=3.2in]{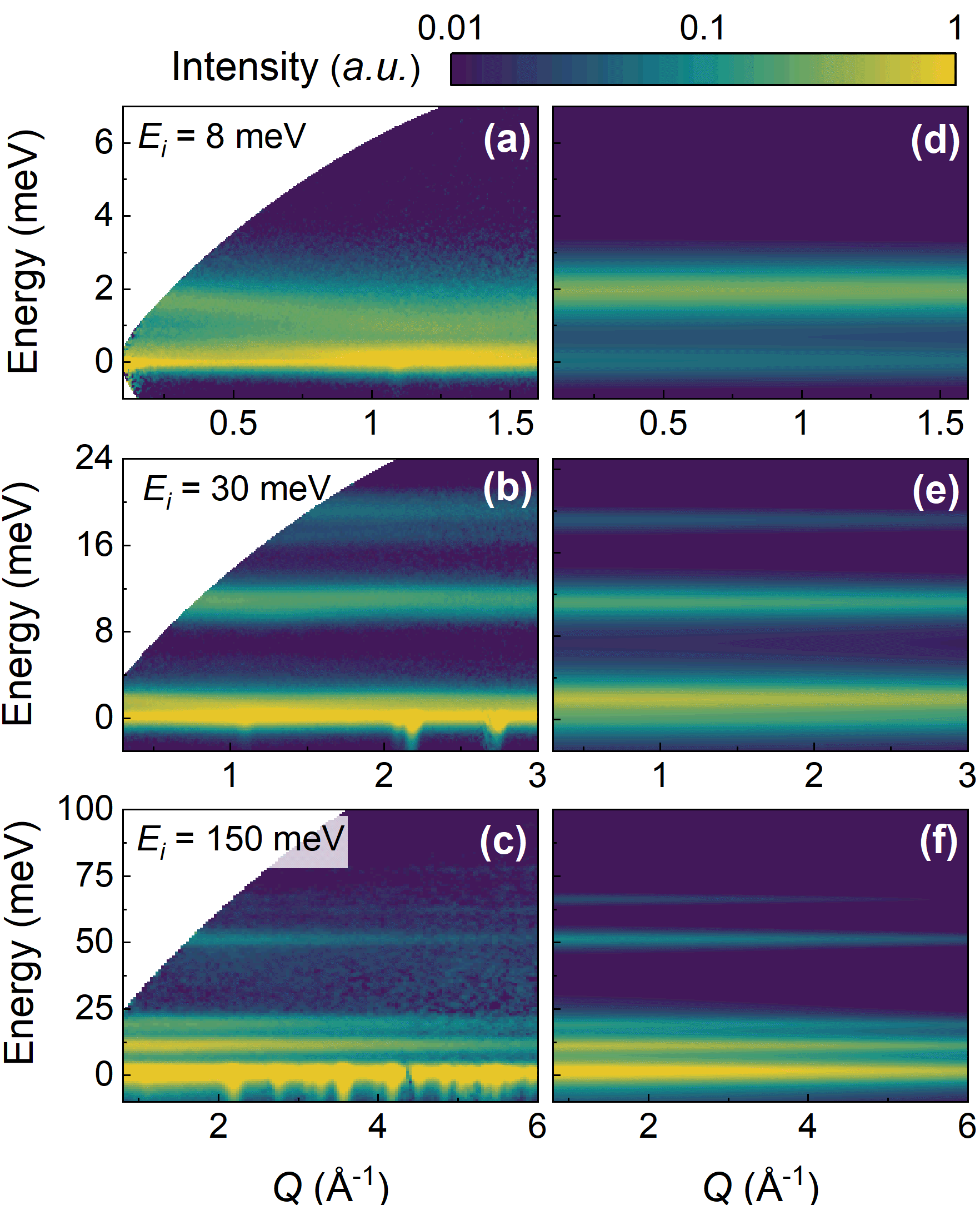}
\par
\caption{Inelastic neutron scattering spectra of the crystal electric field excitations in Tb$_2$Ge$_2$O$_7$ measured at $T = 2$~K with an incident neutron energy of (a) 8~meV, (b) 30~meV, and (c) 150~meV. (d-f) Calculated spectra obtained by refining the crystal electric field Hamiltonian. The calculation shows excellent agreement with the experimental data but does not account for the dispersion of the lowest energy crystal field level, which is due to spin correlations. Note as well that the crystal field level near 20~meV is significantly broadened due to phonon coupling (see text for discussion). An empty can measured under identical instrument configurations has been subtracted from each data set.}
\label{CrystalField}
\end{figure}

The rare earth ion in the $Fd\bar{3}m$ pyrochlore structure sits at the center of an eight-fold coordinate oxygen environment with point group symmetry $D_{3d}$. The crystal field environment
splits the $(2{\text{J}}+1)=13$-fold degeneracy of the spin-orbit ground state manifold for Tb$^{3+}$ into five singlets and four non-Kramers doublets. In order to determine the single ion crystal field states in Tb$_2$Ge$_2$O$_7$, we probed the transitions between them using inelastic neutron scattering. The results of these measurements at three different incident energies ($E_{\rm i} = 8$, 30, and 150~meV) at $T=2$~K are presented in Figs.~\ref{CrystalField}(a-c). The first excited crystal field state in Tb$_2$Ge$_2$O$_7$ sits just $1.5$~meV above the ground state, and is thus not well-separated from it. While crystal field excitations are typically dispersionless, this low-lying crystal field picks up significant Tb-Tb interaction-induced dispersion due to its proximity to the ground state~\cite{kao2003understanding}. Pronounced dispersion is also found in the low-lying crystal fields of both Tb$_2$Ti$_2$O$_7$~\cite{zhang2014neutron,princep2015crystal,ruminy2016crystal} and Tb$_2$Sn$_2$O$_7$~\cite{mirebeau2007magnetic,zhang2014neutron}, as well as in the Er$_2B_2$O$_7$ ($B=$~Ge, Ti, Pt, and Sn) pyrochlores where the first excited crystal field is separated from the ground state by approximately 6~meV~\cite{gaudet2018effect,rau2016order}. Prominent crystal field excitations for Tb$_2$Ge$_2$O$_7$ are also observed at 11, 20, and 50~meV (Figs.~\ref{CrystalField}(b-c)).

\begin{table}[t]
\caption{Comparison of the calculated and experimentally determined crystal field scheme (energies, $E$, and relative intensities, $I$) for Tb$^{3+}$ in Tb$_2$Ge$_2$O$_7$ ($\chi^2$= 4.1). The relative intensities are normalized to the second excited state. Doublets and singlets are referred to as D and S, respectively. For the states where $E_{\rm obs}$ and $I_{{\rm obs}}$ are marked by $-$, the experimental intensities were too small to observe the transition.}
\begin{tabular}{lC{1.4cm}cC{1.7cm}cC{1.7cm}}
\hline
\hline
\multicolumn{2}{c}{ } & \begin{tabular}[c]{@{}c@{}}$E_{\text{obs}}$\\ (meV)\end{tabular} & \begin{tabular}[c]{@{}c@{}}$E_{\text{calc}}$\\ (meV)\end{tabular} & \begin{tabular}[c]{@{}c@{}}$I_{\text{obs}}$\\ (a.u.)\end{tabular} & \begin{tabular}[c]{@{}c@{}}$I_{\text{calc}}$\\ (a.u.)\end{tabular} \\ \hline
GS & D & 0 & 0 & & \\
1 & D & 1.5(3) & 1.5 & 1.7(3) & 2.8 \\
2 & S & 11.0(3) & 11.6 & 1 & 1 \\
\multirow{2}{*}{3$^a$} & \multirow{2}{*}{S} & 16.9(3) & \multirow{2}{*}{18.9} & 0.05(2) & \multirow{2}{*}{0.14} \\
 & & 19.2(3) & & 0.31(3) & \\
4 & D & $-$ & 49.0 & $-$ & 0.002\\
5 & S & $-$ & 49.7 & $-$ & 0.002\\
6 & S & 51.1(5) & 51.1 & 0.12(5) & 0.22 \\
7 & D & 66.3(6) & 66.3 & 0.02(2) & 0.07 \\
8 & S & $-$ & 73.1 & $-$ & 0.003\\ \hline
\hline
\end{tabular}
\vspace{1ex}

 {\raggedright $^a$ As described in the text, the 3rd excited state is split by a vibronic bound state. For the purpose of our CF refinement the energy was taken as 19.2(3)~meV and the intensity was taken as the sum, 0.36(5). \par}
\label{CEFenergy}
\end{table}

\begin{table*}[]
\caption{Spectral decomposition of the 13 crystal field eigenstates of Tb$_2$Ge$_2$O$_7$ expressed in the \ket{{\text{J}}=6,{\text{J}}^z} basis states corresponding to the $^7$F$_{6}$ spin-orbit ground state manifold of the Tb$^{3+}$ ions. The labels in the first column are the same as in Table~\ref{CEFenergy} where $\ket{1{\rm st}}$ and $\ket{2{\rm nd}}$ are the eigenstates making up the ground state (GS) doublet.} 
\begin{tabular}{C{0.8cm}C{1.05cm}C{1.05cm}C{1.05cm}C{1.05cm}C{1.05cm}C{1.05cm}C{1.05cm}C{1.05cm}C{1.05cm}C{1.05cm}C{1.05cm}C{1.05cm}C{1.05cm}C{1.05cm}}
\hline
\hline
 & & \ket{-6} & \ket{-5} & \ket{-4} & \ket{-3} & \ket{-2} & \ket{-1} & \ket{0} & \ket{1} & \ket{2} & \ket{3} & \ket{4} & \ket{5} & \ket{6} \\ \hline
\multirow{2}{*}{GS D} & \ket{1{\rm st}} & 0 & 0 & -0.827 & 0 & 0 & -0.103 & 0 & 0 & -0.230 & 0 & 0 & 0.502 & 0 \\ 
& \ket{2{\rm nd}} & 0 & 0.502 & 0 & 0 & 0.230 & 0 & 0 & -0.103 & 0 & 0 & 0.827 & 0 & 0 \\ 
\multirow{2}{*}{1 \;\, D}& \ket{3{\rm rd}} & 0 & 0 & 0.530 & 0 & 0 & 0.145 & 0 & 0 & -0.193 & 0 & 0 & 0.813 & 0 \\ 
& \ket{4{\rm th}} & 0 & -0.813 & 0 & 0 & -0.193 & 0 & 0 & -0.145 & 0 & 0 & 0.530 & 0 & 0 \\ 
2 \;\; S & \ket{5{\rm th}} & -0.226 & 0 & 0 & -0.670 & 0 & 0 & 0 & 0 & 0 & -0.670 & 0 & 0 & 0.226 \\ 
3 \;\; S & \ket{6{\rm th}} & -0.285 & 0 & 0 & -0.634 & 0 & 0 & -0.186 & 0 & 0 & 0.634 & 0 & 0 & -0.285 \\ 
\multirow{2}{*}{4 \;\, D} & \ket{7{\rm th}} & 0 & -0.293 & 0 & 0 & 0.946 & 0 & 0 & 0.125 & 0.00 & 0 & 0.070 & 0 & 0 \\ 
& \ket{8{\rm th}} & 0 & 0 & -0.070 & 0 & 0 & -0.125 & 0 & 0 & 0.946 & 0 & 0 & 0.293 & 0 \\ 
5 \;\; S & \ket{9{\rm th}} & -0.670 & 0 & 0 & 0.226 & 0 & 0 & 0 & 0 & 0 & 0.226 & 0 & 0 & 0.670 \\ 
6 \;\; S & \ket{10{\rm th}} & -0.647 & 0 & 0 & 0.281 & 0 & 0 & 0.066 & 0 & 0 & -0.281 & 0 & 0 & -0.647 \\ 
\multirow{2}{*}{7 \;\, D}& \ket{11{\rm th}} & 0 & 0 & -0.175 & 0 & 0 & -0.976 & 0 & 0 & 0.125 & 0 & 0 & -0.031 & 0 \\ 
& \ket{12{\rm th}} & 0 & 0.031 & 0 & 0 & 0.125 & 0 & 0 & -0.976 & 0 & 0 & -0.175 & 0 & 0 \\ 
8 \;\; S & \ket{13{\rm th}} & 0.010 & 0 & 0 & 0.139 & 0 & 0 & 0.980 & 0 & 0 & -0.139 & 0 & 0 & 0.010 \\ \hline
\hline
\end{tabular}
\label{CEFcomp}
\end{table*}

We have analyzed the crystal field scheme of Tb$_2$Ge$_2$O$_7$ using the same method employed in Ref.~\cite{gaudet2018effect}. In short, the crystal field (CF) Hamiltonian was expressed in terms of Stevens operators, ${\mathcal H}_{\text{cf}} = \sum_{mn} B^n_m \hat{O}^n_{m} ({\text{J}}^z,
{\text{J}}^+,
{\text{J}}^-)$,
 and diagonalized within the $(2{\text{J}}+1)=13$ spin-orbit states of the $^7$F$_6$ ground state of the Tb$^{3+}$ ion. The six adjustable parameters in ${\mathcal H}_{\text{cf}}$ were determined via $\chi^2$ minimization against the experimentally observed energies and relative scattered intensities of the crystal field excitations observed in Figs.~\ref{CrystalField}(a-c). The best agreement with our data was obtained with $B^0_2=-0.27$~meV, $B^0_4=0.0051$~meV, $B^3_4=0.0459$~meV, $B^0_6=-1.46\cdot10^{-6}$~meV, $B^3_6 = 1.06 \cdot10^{-4}$~meV and $B^6_6 = -1.75\cdot10^{-4}$~meV. The observed and calculated energies and relative scattered intensities of the crystal field excitations are reported in Table~\ref{CEFenergy}. The calculated neutron spectra is shown side-by-side with the experimental data in Figs.~\ref{CrystalField}(d-f), revealing excellent agreement. For temperatures above $10$~K, where correlation effects are negligible, the free-ion susceptibility computed with our refined CF Hamiltonian also agrees well with the previously measured susceptibility of Tb$_2$Ge$_2$O$_7$~\cite{hallas2014incipient}.

The detailed composition of all the crystal field eigenfunctions are reported in Table~\ref{CEFcomp}. The crystal field energy scheme determined for Tb$_2$Ge$_2$O$_7$ is similar to those of Tb$_2$Ti$_2$O$_7$ and Tb$_2$Sn$_2$O$_7$. For all three compounds, two crystal field doublets are found below 2~meV, which are followed by two singlet states between 10 and 20~meV. All other excited crystal field levels are located above 40~meV. In these three terbium pyrochlores, the ground state and first excited state are composed predominantly of the \ket{{\text{J}}^z=\pm5} and \ket{{\text{J}}^z=\pm4} states. In both Tb$_2$Ge$_2$O$_7$ and Tb$_2$Ti$_2$O$_7$, the ground state doublet is primarily \ket{{\text{J}}^z=\pm4} while the first excited doublet is dominated by the \ket{{\text{J}}^z=\pm5} states. The sequence is inverted in the case of Tb$_2$Sn$_2$O$_7$ with \ket{{\text{J}}^z=\pm5} forming the ground state doublet and \ket{{\text{J}}^z=\pm4} the first excited crystal field level. In all three compounds, the transverse moment of the crystal field ground doublet is strictly zero~\cite{onoda2011quantumspinice,lee2012generic,rau2019frustrated}
and the Ising moment is on the order of 5~$\mu_{\rm B}$ for Tb$_2$Ti$_2$O$_7$ and Tb$_2$Sn$_2$O$_7$, while we refine a significantly smaller moment of 2.1~$\mu_{\rm B}$ for Tb$_2$Ge$_2$O$_7$~\cite{CFmoment}.

The calculated crystal field scheme of Tb$_2$Ge$_2$O$_7$ predicts an energy level at 18.9~meV, which is well-separated from all other excited states, and should be visible as a single transition from the ground state doublet to the $\ket{6{\text{th}}}$ excited singlet state (see Table~\ref{CEFcomp}). However, as can be seen in Fig.~\ref{CrystalField}(b), our experiment at 2~K shows two closely spaced excitations that are centered at 16.9(3) and 19.2(3)~meV. Our attempts to refine a crystal field Hamiltonian that treats both as excitations out of the crystal field ground state yielded solutions that were clearly inconsistent with the experimental data. Furthermore, only one crystal field level is expected in this energy range based on a scaling approximation of the crystal field $B^n_m$~\cite{gaudet2018effect,bertin2012crystalfield}. An additional transition involving the same $\ket{6{\text{th}}}$ excited singlet state has also been observed in Tb$_2$Ti$_2$O$_7$~\cite{zhang2014neutron,princep2015crystal,ruminy2016crystal}, but not in Tb$_2$Sn$_2$O$_7$~\cite{mirebeau2007magnetic,zhang2014neutron}. 

As has been proposed for Tb$_2$Ti$_2$O$_7$, we suggest that the additional excitation at 16.9(3)~meV in Tb$_2$Ge$_2$O$_7$ is produced by a \textit{vibronic bound state}, which is the hybridization of an optical phonon mode with a crystal field state excitation. A vibronic bound state has also been observed and modeled in holmium pyrochlores, though there it involves a phonon and an excited doublet as opposed to an excited singlet in the present case~\cite{Gaudet2018vibron}. Within this scenario, the neutron scattering cross-section of an optical phonon acquires a magnetic form factor that arises from admixing with the nearby CF state. This admixing, which is mediated by magneto-elastic interactions, is only allowed when the phonon involved is of the same symmetry as that of the quadrupolar operators characterizing the local distortion of the Tb$^{3+}$ $D_{3d}$ point group symmetry. 

In the case of Tb$_2$Ti$_2$O$_7$, symmetry analysis reveals that the quadrupolar operators, $\hat{O}_{yz} = {\text{J}}^z{\text{J}}^y+{\text{J}}^y{\text{J}}^z$ and $\hat{O}_{xz} = {\text{J}}^z{\text{J}}^x+{\text{J}}^x{\text{J}}^z$, can indeed admix an optical phonon and the ``bare'' $\ket{6{\text{th}}}$ excited crystal field state which would, otherwise free of admixing with the phonon, give a single observable transition from the ground state at 2~K~\cite{ruminy2016crystal}. The strength of this coupling is proportional to the matrix element of the quadrupolar operators $\sum_{i}\left(|\bra{v_f}\hat{O}{_{yz}}\ket{v_i}|^2+|\bra{v_f}\hat{O}{_{xz}}\ket{v_i}|^2\right)$ where $v_f$ ($\ket{6{\text{th}}}$) and $v_i$ ($\ket{1{\text{st}}}$ and $\ket{2{\text{nd}}}$) are the final and initial crystal field states involved in the excitation, respectively. The matrix element of the quadrupolar operators is stronger for a ground state composed of \ket{{\text {J}}^z =\pm 4} states, as is the case for Tb$_2$Ge$_2$O$_7$ and Tb$_2$Ti$_2$O$_7$, as compared to a ground state composed of \ket{{\text J}^z=\pm 5} states, as in Tb$_2$Sn$_2$O$_7$. Indeed, the calculated matrix elements for the vibronic bound state in Tb$_2$Ti$_2$O$_7$ and Tb$_2$Ge$_2$O$_7$ are almost two orders of magnitude larger than for Tb$_2$Sn$_2$O$_7$, providing a natural explanation for why it is not observed in the latter.


\subsection{Magnetic Structure Determination}
\label{sec:magstrucdeterm}

\begin{figure}[tbp]
\linespread{1}
\par
\includegraphics[width=3.2in]{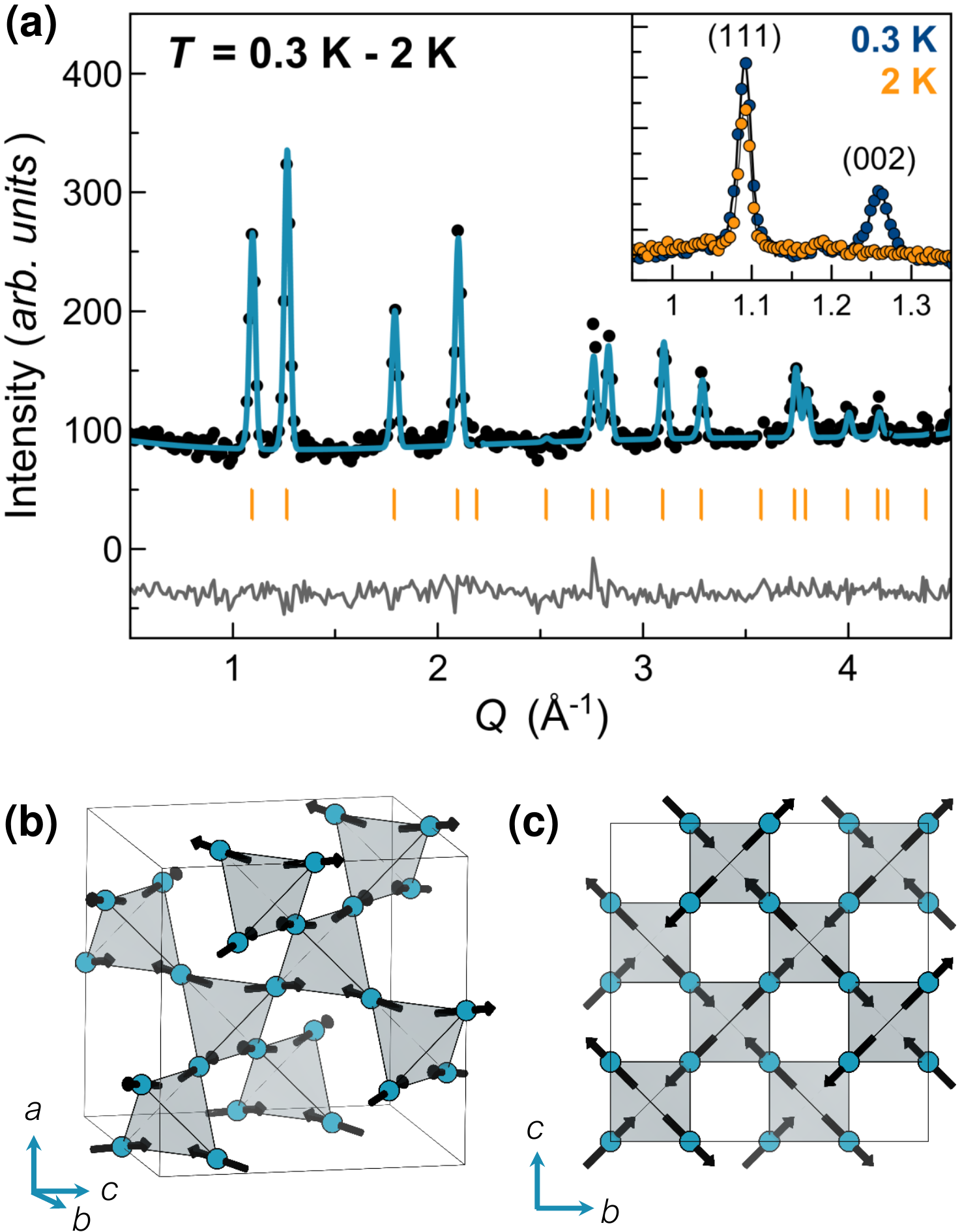}
\par
\caption{(a) The magnetic structure of Tb$_2$Ge$_2$O$_7$ in the intermediate regime, $T_c < T < T^*$, was determined by a Rietveld refinement of the powder neutron diffraction pattern at $T=0.3$~K, where the magnetic pattern was isolated by subtracting a 2~K background (shown in the inset). The best refinement was obtained with the $\mathbf{k}={\mathbf {0}}$, $\Gamma_9$ irreducible representation ($\chi^2 = 1.83$), giving an ordered moment of 1.95(1)~$\mu_{\rm B}$/Tb$^{3+}$. (b,c) This splayed ferromagnetic structure is a canted version of spin ice; each tetrahedron obeys two-in, two-out ice rule for the $\langle 111 \rangle$ components of the magnetic moments, while the whole moment is canted $\alpha=24.5^{\circ}$ away from the local $\langle$111$\rangle$.}
\label{Refinements}
\end{figure}

In order to expose the complex phase behavior in Tb$_2$Ge$_2$O$_7$ hinted by its heat capacity (see Fig.~\ref{Tb_Cp}), we begin by interrogating its static magnetic properties via neutron diffraction measurements. We first consider the regime that is below the heat capacity anomaly at $T^*=1.1$~K, but above the first-order transition at $T_c=0.25$~K (Fig.~\ref{Tb_Cp}). Upon cooling through $T^* = 1.1$~K, we observe the formation of magnetic Bragg reflections, as observed for the (111) and (002) positions in the inset of Fig.~\ref{Refinements}(a). The magnetic diffraction pattern is isolated by subtracting the $T=2$~K data set from the $T=0.3$~K data set, the result of which is shown in the main panel of Fig.~\ref{Refinements}(a). All of the magnetic Bragg peaks at $T=0.3$~K can be indexed with a $\mathbf{k}={\mathbf {0}}$ propagation vector relative to the $Fd\bar{3}m$ space group symmetry for which there are four allowed irreducible representations for the $16d$ Wyckoff site: $\Gamma_3$, $\Gamma_5$, $\Gamma_7$, and $\Gamma_9$. Rietveld refinements were attempted with each of these magnetic structures and only $\Gamma_9$ was able to capture all of the observed magnetic reflections, yielding excellent agreement with the experimental data (Fig.~\ref{Refinements}(a)). 

\begin{figure}[tbp]
\linespread{1}
\par
\includegraphics[width=3.2in]{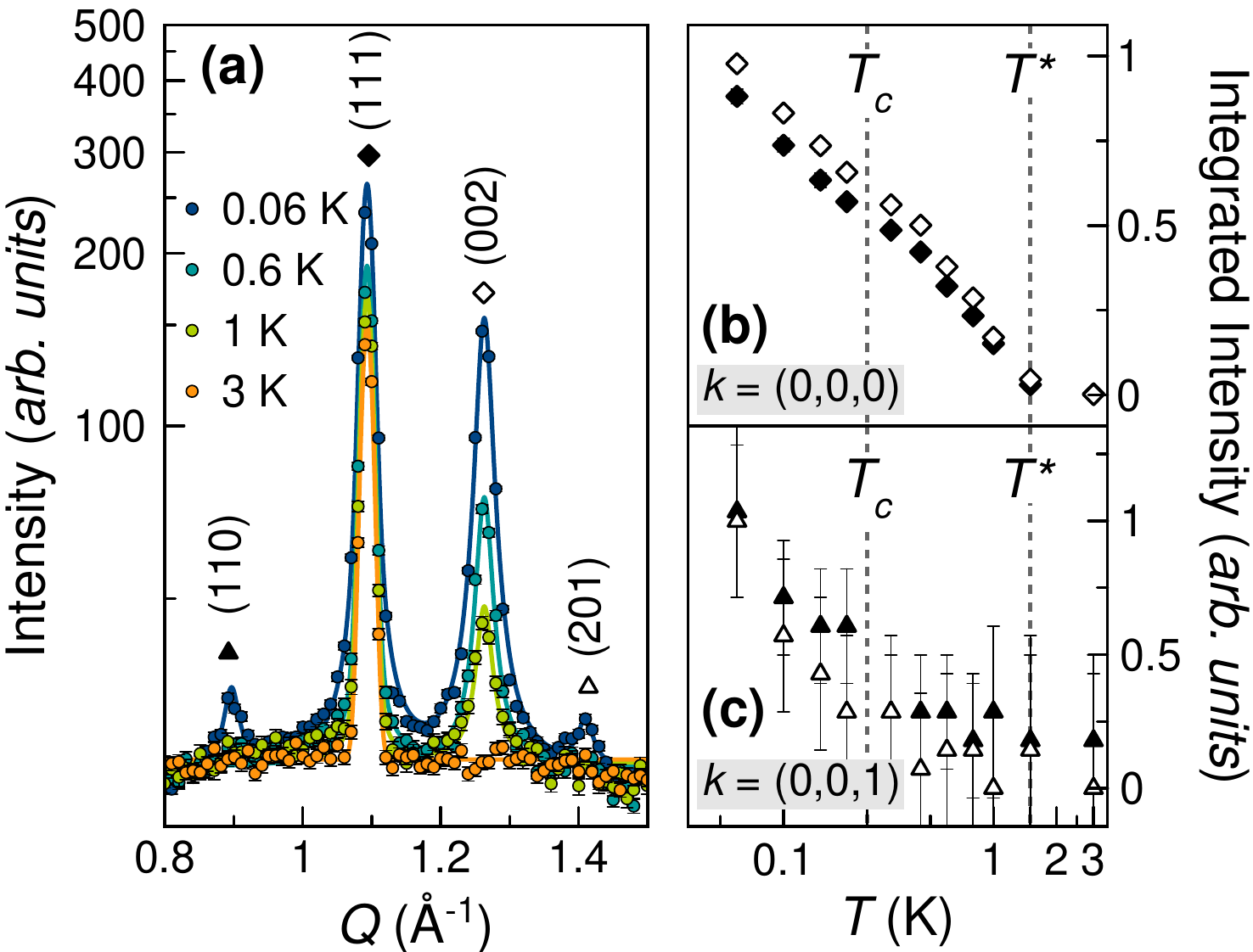}
\par
\caption{(a) The diffraction pattern of Tb$_2$Ge$_2$O$_7$ down to $T=0.06$~K was isolated by integrating over the elastic channel of the time-of-flight data, $E = $~[$-0.05$, 0.05]~meV. The $\mathbf{k}={\mathbf {0}}$ magnetic Bragg peaks that form below $T^*=1.1$~K continue to grow below $T_c = 0.25$~K. Two new magnetic Bragg peaks form at (110) and (201), consistent with a $\mathbf{k}=(0,0,1)$ antiferromagnetic distortion of this structure. The order parameters for the (b) $\mathbf{k}={\mathbf {0}}$ and (c) $\mathbf{k}=(0,0,1)$ structures are obtained by integrating over the four Bragg peaks seen in panel (a) that are labelled by the filled and open diamonds for (111) and (002) (for panel (b))
and the filled and open triangles for (110) and (201) (for panel (c)), respectively. The $\mathbf{k}={\mathbf {0}}$ order begins to develop at $T^*=1.1$~K and then sharply increases below $T_c=0.25$~K, while the $\mathbf{k}=(0,0,1)$ order onsets at $T_c=0.25$~K.}
\label{OrderParameter}
\end{figure}

The $\Gamma_9$ irreducible representation is made up of two basis vectors, which produce ordered states related to the two-in, two-out spin ice states. In the first of these, the spins are ferromagnetically aligned along the cubic $\langle$100$\rangle$ directions while in the second, pairs of anti-aligned spins point along $\langle$110$\rangle$ directions, which are the Tb-Tb bond axes \footnote{The ordered phases whose order parameters corresponding to $\Gamma_9$ irreducible representation are superpositions of colinear FM and noncolinear FM, which are defined in Ref.~\cite{yan2017theory}. Note the order parameters of the $\Gamma_9$ phase in Ref.~\cite{yan2017theory}, denoting as ${\bm m}_{\rm T_{1,A}}$ and ${\bm m}_{\rm T_{1,B}}$ are linear combinations of the $\Gamma_9$ order parameter in Ref.~\cite{rau2019frustrated}, denoting as ${\bm M}_{T_{1g}}$ and ${\bm M}_{T_{1'g}}$, which we adopt in this work. The $\Gamma_9$ order parameters in Table~\ref{tab:orders} ${\bm m}_{\Gamma_{9-\rm SI}}$ and ${\bm m}_{\Gamma_{9-\rm SFM}}$ are correspond to ${\bm M}_{T_{1g}}$ and ${\bm M}_{T_{1'g}}$ in Ref.~\cite{rau2019frustrated}, respectively.}. At $T=0.3$~K, the linear combination of basis vectors that describes Tb$_2$Ge$_2$O$_7$ is $\psi_{\langle100\rangle} - 0.65 \cdot \psi_{\langle110\rangle}$, yielding the canted spin ice state shown in Fig.~\ref{Refinements}(b). As can be seen by looking along the $a$ axis, as in Fig.~\ref{Refinements}(c), there is indeed a two-in, two-out spin ice component to this ordered state. However, the moments are significantly canted from the Ising axis with each moment making an angle $\alpha=24.5^{\circ}$ with the local $\langle$111$\rangle$ axis. This result is initially surprising, as our crystal field analysis revealed that the ground state doublet magnetic moment of Tb$^{3+}$ is strictly Ising-like, and hence, in the absence of other factors, the magnetic moments should be confined to the $\langle$111$\rangle$ axes. This is the first piece of evidence for the strong influence of the low-lying crystal field level at 1.5~meV. As has been previously shown, admixture between the ground state and the first excited state can disrupt the Ising anisotropy of the ground state doublet~\cite{molavian2007dynamically,mcclarty2010soft}, resulting in a canted spin ice state where the spins are strongly canted towards the $\langle 110 \rangle$ directions (the moments are canted only by $\beta = 10.9^{\circ}$ away from $\langle$110$\rangle$). The ordered magnetic moment at $T = 0.3$~K is 1.95(1)~$\mu_{\rm B}$ per Tb$^{3+}$, which is almost the full magnetic moment assigned to the single ion crystal field doublet given in Sec.~\ref{subsec:cf-analysis}, and there is a net ferromagnetic moment of 0.68~$\mu_{\rm B}$ per tetrahedron (0.17~$\mu_{\rm B}$/Tb$^{3+}$) along $\langle 100 \rangle$.

\begin{figure*}[tbp]
\linespread{1}
\par
\includegraphics[width=7in]{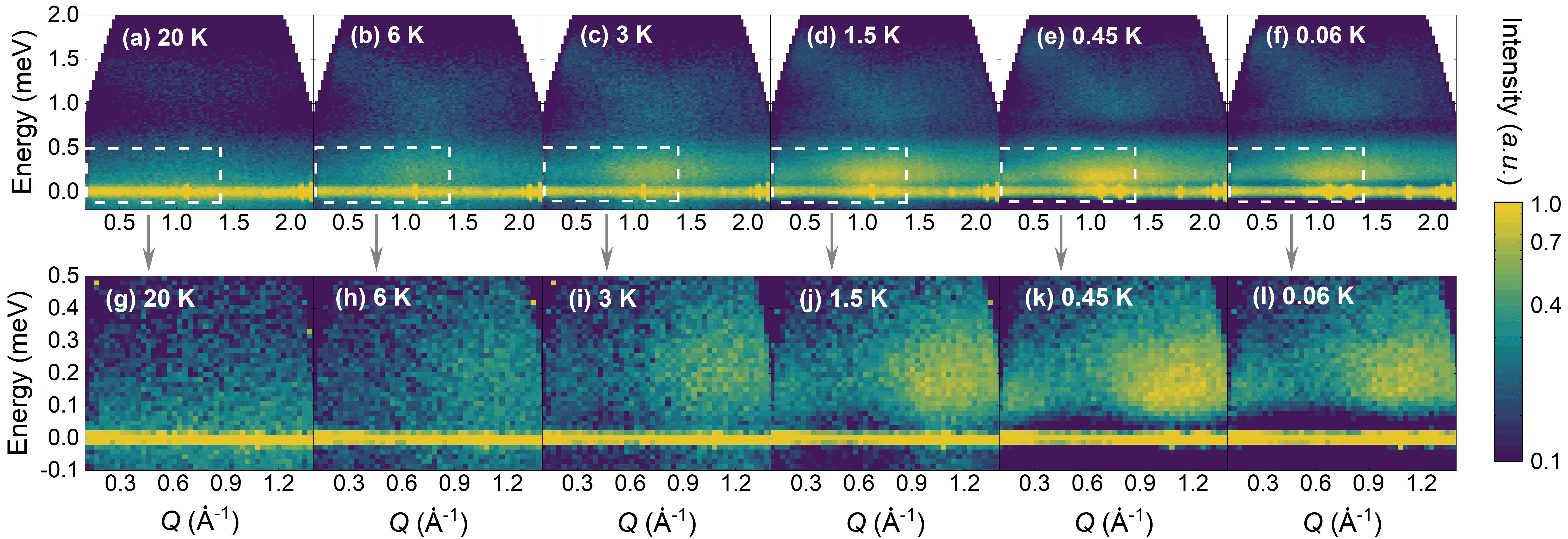}
\par
\caption{The temperature dependence of the low energy spin excitations of Tb$_2$Ge$_2$O$_7$. The upper and lower panels show inelastic scattering at the same temperatures with two different incident energies, (a-f) $E_{\rm i} = 3.3$~meV and (g-l) $E_{\rm i}=1.3$~meV, respectively. A low energy spin excitation, centered around $Q=1.1$~\AA$^{-1}$, forms below 6~K. This excitation develops an $E_{\text{gap}} = 0.18$~meV gap at the lowest temperatures. An additional lobe of scattering accumulates at low $Q$, as can be clearly seen most clearly in panel (k). An empty can measured under identical instrument configurations has been subtracted from each data set.}
\label{DCS_Slices}
\end{figure*}

Upon examination of the inset of Fig.~\ref{Refinements}(a), one can see that the widths of the magnetic peaks at (111) and (002) are broadened as compared to the (111) nuclear peak at $T = 2$~K and are therefore not limited by instrumental resolution. This broadening signifies that while the magnetic order has a rather long correlation length, it remains significantly shorter than the system size. We quantify the correlation length by fitting these Bragg peaks to a Lorentzian line shape:
\begin{equation}
L(Q) = \frac{1}{\pi} \cdot \frac{\kappa}{(Q-Q_0)^2 + \kappa^2},
\end{equation}
where $Q_0$ is the peak center and $\kappa$ is the half width at half maximum, which is inversely related to the mean correlation length, $\xi \propto 1/\kappa$. The (111) nuclear Bragg peak at $T = 2$~K is assumed to be resolution-limited. The (002) magnetic Bragg peak displays a width roughly double that of the resolution limit, giving a correlation length of $\xi = 67 \pm 2$~\AA, approximately seven conventional cubic unit cells.

Next, we study the evolution of Tb$_2$Ge$_2$O$_7$'s magnetic state at temperatures below its $T_c = 0.25$~K first-order transition. In this regime, we observe that the magnetic Bragg peaks associated with the $\mathbf{k}={\mathbf {0}}$ $\Gamma_9$ order that first formed at $T^* =1.1$~K continue to intensify, signifying an increase in the ordered moment (Fig.~\ref{OrderParameter}(a,b)). Interestingly, the peak widths do not become narrower and thus, the correlation length for the magnetic order remains essentially unchanged. Below $T_c=0.25$~K, we begin to also observe the formation of two new Bragg reflections at $Q = 0.91$~\AA$^{-1}$~and $Q=1.41$~\AA$^{-1}$ (Fig.~\ref{OrderParameter}(a,c)). Neither of these positions are allowed by the selection rules for the $Fd\bar{3}m$ pyrochlore structure; they can, however, be indexed as (110) and (201) respectively, with a $\mathbf{k}=(0,0,1)$ propagation vector. Such a phase transition reduces the order of the space group by three and cannot occur in a continuous fashion~\cite{wills2000new} and is thus consistent with the first-order-like transition we observe in the heat capacity measurements (see Fig.~\ref{Tb_Cp}). This symmetry lowering transition could be accompanied by a tetragonal distortion of the crystal lattice, but this is not required. The $Q$ resolution in our measurement does not allow us to definitively comment on whether this magnetic transition is accompanied by a structural transition.

In contrast to a $\mathbf{k}={\mathbf {0}}$ order, a $\mathbf{k}=(0,0,1)$ propagation vector indicates that the magnetic order breaks the face-centered cubic selection rules of the underlying $Fd\bar{3}m$ nuclear structure such that the spin configuration on the four tetrahedra per conventional cubic unit cell are no longer identical. Two irreducible representations within $\mathbf{k}=(0,0,1)$, $\Gamma_2$ ($\psi_4$ and $\psi_6$ only) and $\Gamma_3$ ($\psi_7$ and $\psi_8$), can reasonably account for the additional Bragg peaks that form below $T_c = 0.25$~K~\footnote{The ordered state of classical dipolar Ising spin ice~\cite{melko2001DSILRO,melko2004DSIMC} with ordering wave vector ${\mathbf{k}} = (0,0,1)$ is obtained through linear combinations within $\Gamma_2$ that also involve $\psi_3$ and $\psi_5$. However, this state is inconsistent with the ${\mathbf{k}} = (0,0,1)$ order in Tb$_2$Ge$_2$O$_7$ as it produces an intense magnetic Bragg peak at (001) that is absent in our data.}. Both of these structures are antiferromagnetic; in the first one, the spins point along the crystallographic $c$-axis while in the second the spins lie in the local XY plane. However, with the present data we are not able to uniquely distinguish between these two scenarios. It is worth emphasizing that these additional peaks are of very small intensity as compared to the primary $\mathbf{k}={\mathbf {0}}$ peaks (note the log intensity scale in Fig.~\ref{OrderParameter}), reflecting the fact that the majority of the ordered moment remains in the canted spin ice state illustrated in Fig.~\ref{Refinements}(b,c). At $T=0.06$~K, the canted spin ice moment has increased to 2.46(1)~$\mu_{\rm B}$ per Tb$^{3+}$ but without us being to resolve a change in the canting angle. At the same temperature, the antiferromagnetic $\mathbf{k}=(0,0,1)$ ordered moment is approximately 0.5~$\mu_{\rm B}$ per Tb$^{3+}$, as estimated by Rietveld refinements with either the $\Gamma_2$ or the $\Gamma_3$ irreducible representation.


\subsection{Inelastic Scattering from Collective Spin Excitations}
\label{subsec:INS}

We already noted above two peculiarities regarding the development of the magnetic correlations in Tb$_2$Ge$_2$O$_7$. First, these proceed through two well-separated temperature scales, $T^*$ and $T_c$, and it is not immediately clear from the specific heat or the neutron diffraction measurements what is the nature of the strongly correlated state within the $(T_c < T < T^*)$ temperature window. Secondly, the large canting of the dipole moments away from their local $\langle 111\rangle$ Ising axes (at $T < T^*$) speaks to the important role that interaction-induced admixing of the two lowest doublets plays in the magnetic correlations of this material. Our low-energy inelastic neutron scattering measurements on Tb$_2$Ge$_2$O$_7$, presented in Fig.~\ref{DCS_Slices}, shed light on these two issues.

Two data sets were collected at each temperature; one with an incident energy $E_{\rm i} = 3.3$~meV (top panels) and the other with a smaller incident energy $E_{\rm i} = 1.3$~meV (bottom panels). The first feature we describe is visible only in the top set of panels -- a dispersive mode centered around 1.5~meV. This low-lying crystal field excitation, which picks up dispersion from ion-ion interactions~\cite{kao2003understanding}, is the same one that we observed previously in the higher incident energy inelastic neutron scattering measurements that were used to determine the single ion properties (Fig.~\ref{CrystalField}(a)). This feature intensifies upon cooling through the Schottky anomaly centered at $T_{\Delta} = 6$ K as it is proportional to the Tb$^{3+}$ single ion density of states for the crystal field ground state.

A second inelastic feature, centered near 0.18~meV and 1.1~\AA$^{-1}$ at the lowest temperatures, is visible in both sets of panels (Fig.~\ref{DCS_Slices}). This low energy mode, a collective magnetic excitation of some sort, has an unusual temperature dependence. Indeed, this excitation begins to form near $T_{\Delta} = 6$~K, and grows sharper and more intense all the way down to $T = 0.45$~K (Fig.~\ref{DCS_Slices}(k)). Below $T_c = 0.25$~K, this excitation appears to become fully gapped and begins to decrease in intensity. We define the energy gap, $E_{\text{gap}} = 0.18$~meV, of this mode as the energy difference between the elastic line and the mode's maximum intensity at the lowest measured temperature, $T=0.05$~K. Lastly, beginning near $T^* = 1.1$~K and as shown in Fig.~\ref{DCS_Slices}(j), a second lobe of scattering can be resolved at low $Q$ below 0.3~\AA$^{-1}$. This latter scattering, which is likely centered at $Q=0$~\AA$^{-1}$, is consistent with the net ferromagnetic polarization of the ordered $\Gamma_9$ magnetic structure, as well as the short-range ferromagnetic scattering previously measured with polarized neutron diffraction~\cite{hallas2014incipient}.

\begin{figure}[tbp]
\linespread{1}
\par
\includegraphics[width=3.2in]{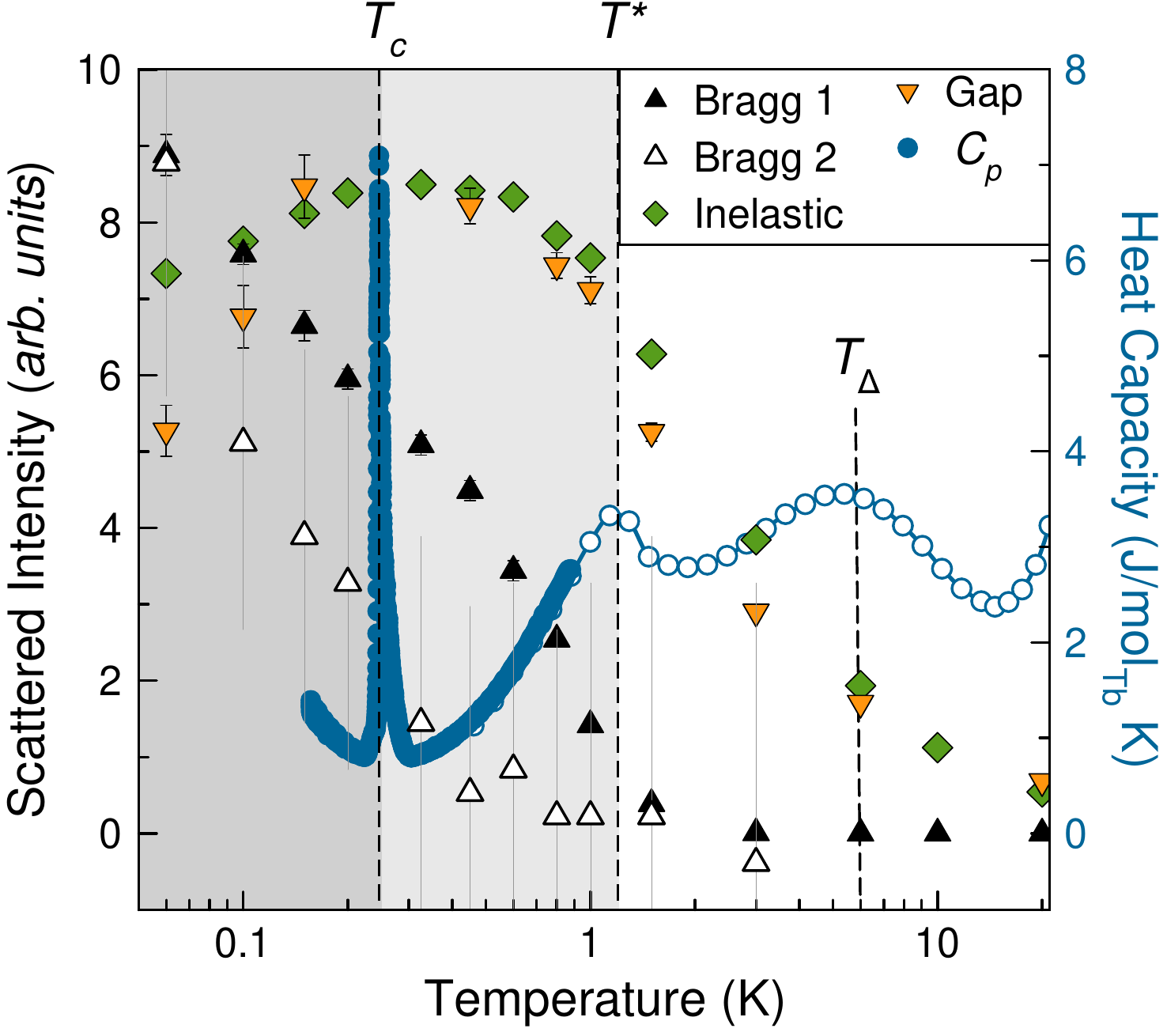}
\par
\caption{Compilation of the heat capacity of Tb$_2$Ge$_2$O$_7$,
$C_P$ plotted alongside the elastic, inelastic and in-gap scattered neutron intensity. The $\mathbf{k}={\mathbf {0}}$ and $\mathbf{k}=(0,0,1)$ elastic scattering are the integrated intensities of the (002) and (110) magnetic Bragg peaks, filled and open black triangles, respectively. The inelastic scattering is obtained by integrating between $Q = 0.6-1.7$~\AA$^{-1}$ and $E = 0.1-0.4$~meV in the $E_{\rm i} = 3.3$~meV data set and divided out by the Bose factor. The in-gap scattering (filled yellow triangles) is obtained by integrating between $Q = 0.2-1.3$~\AA$^{-1}$ and $E = 0.025-0.05$~meV in the $E_{\rm i} = 1.3$~meV data set and correcting for the Bose factor. Bragg elastic scattering onsets at $T^*=1.1$~K and rapidly rises below $T_c=0.25$~K with increasing upwards curvature. The rate of increase of the spectral weight originating from the collective spin excitation (green diamonds) diminishes from $T^*$ down to $T_c$. Below $T_{c}$ the intensity of the collective mode decreases as the gap hardens. 
}
\label{fig:Summary}
\end{figure}

The phase behavior of Tb$_2$Ge$_2$O$_7$ is intriguing and unconventional and is readily appreciated by considering Fig.~\ref{fig:Summary}, where we juxtapose the neutron scattering results in relationship with the three thermodynamic anomalies. The heat capacity data, $C_p$, which is reproduced from Fig.~\ref{Tb_Cp}, is given by the blue circles. Next, the temperature dependence of the $\mathbf{k}={\mathbf {0}}$ and the $\mathbf{k}=(0,0,1)$ order parameters, obtained from the integrated intensity of the (002) and the (110) magnetic Bragg peaks as featured in Fig.~\ref{OrderParameter}(b,c), are given by the black (``Bragg 1'') and white (``Bragg 2'') triangles, respectively. The intensity of the collective excitation at 0.18~meV, represented by the green diamonds, is obtained by integrating the inelastic signal between $Q = 0.6-1.7$~\AA$^{-1}$ and $E = 0.1-0.4$~meV in the $E_{\rm i} = 3.3$~meV data set. Finally, the spectral weight within the gap, represented by the yellow triangles, is obtained by integrating between $Q \in [0.2-1.3]$~\AA$^{-1}$ and $E \in [0.025-0.05]$~meV in the $E_{\rm i} = 1.3$~meV data set. Both the inelastic and gap integrated intensities have been corrected by dividing out the Bose factor such that the quantity plotted is proportional to the imaginary part of the dynamical susceptibility, $\chi''(Q,\omega)$. 
The total scattering intensity for each of these integrations has been independently and arbitrarily scaled and thus, no conclusions should be drawn on the basis of their relative magnitudes.

Using Fig.~\ref{fig:Summary} as guide, we can summarize the evolution of Tb$_2$Ge$_2$O$_7$'s static and dynamic properties as it passes through each of its three thermodynamic anomalies signaled by the three peaks in the heat capacity, $C_p$ (at $T_\Delta$, $T^*$ and $T_c$) upon cooling. At temperatures around $T_{\Delta} = 6$~K, Tb$_2$Ge$_2$O$_7$ crosses over from a thermal paramagnet to a strongly correlated (collective) paramagnet~\cite{villaininsulating1979}. In this temperature regime, the first excited crystal field doublet becomes progressively thermally depopulated resulting in the Schottky-like anomaly in $C_p$ at $T_\Delta$. As a result, crystal field transitions out of that excited doublet state disappear and, simultaneously, a collective spin excitation begins to develop at 0.18 meV. The rapid onset of the elastic component of the scattering occurs upon cooling through $T^* = 1.1$~K, where non-resolution limited Bragg peaks form. Here, Tb$_2$Ge$_2$O$_7$ enters a $\Gamma_9$ state with rather extended correlations, characterized by a correlation length $\xi = 67 \pm 2${\AA}, as described in Sec.~\ref{sec:magstrucdeterm}. Between $T^*$ and $T_c$, the intensity of the $0.18$~meV collective spin excitation (filled green diamonds in Fig.~\ref{fig:Summary}) plateaus, perhaps slightly decreasing just before reaching $T_c$. We note, however, that there exists significant inelastic ``in-gap'' intensity (filled yellow triangles) from $T_\Delta$ down to $T_c$. Finally, $T_c = 0.25$~K marks an apparently first-order transition into a multi-${\mathbf k}$ state where there is a sharp increase in the magnetic Bragg scattering associated with the ordered canted spin ice state that had formed at $T^*$, but now co-existing with an additional antiferromagnetic $\mathbf{k}=(0,0,1)$ component to the ordered moment. Remarkably, the order parameter for this transition remains unsaturated down to $T=0.06$~K~\footnote{Thermal equilibration of our powder sample below $T_c$ cannot be guaranteed hence the nominal sample temperature may deviates from the temperature recorded by the thermometer. This may influence the details of the order parameter plots below $T_c$.}. In this lowest temperature state, a clear spin gap opens as the spectral weight below $E_{\text{gap}} = 0.18$~meV becomes depleted while the intensity of this collective spin excitation also decreases slightly.

With these results in hand, we argue that understanding the complex phase behavior of Tb$_2$Ge$_2$O$_7$ hinges, in large part, on being able to explain the origin of the intense inelastic mode at 0.18~meV that develops below $T_\Delta = 6~{\rm K}$. One might naturally assume that the 0.18 meV mode originates from the collective magnetic excitations within the correlated $\Gamma_9$ spin ice like domains. However, such an interpretation would be too hasty. The origin of this inelastic mode is, as we discuss in Section~\ref{sec:theory}, nontrivial and a full understanding of the mechanism behind it would, we believe, ultimately unravel both the low temperature physics of Tb$_2$Ge$_2$O$_7$ and that of the other members of the terbium pyrochlore family. As a first step in this program, we now discuss a minimal model aimed at capturing the salient features of the inelastic neutron scattering excitations of Tb$_2$Ge$_2$O$_7$ within a state with $\Gamma_9$ correlations.

\section{Theoretical Modeling}
\label{sec:theory}

\subsection{Theoretical Context}
\label{subsec:perspective-modeling}

In most rare earth pyrochlores, the energy gap, $\Delta$, between the crystal field ground state doublet and the first excited states is at least two orders of magnitude larger than any of the interactions between the rare earth ion's angular momenta~\cite{bertin2012crystalfield,gardner2010magnetic,hallas2018experimental,rau2019frustrated}.
In such cases, a nearest neighbor pseudospin-$\sfrac{1}{2}$ Hamiltonian can be used to describe the interactions
\begin{equation}
 {\cal H}_{\text{eff}} = \sum_{\avg{i,j}} J_{ij}^{\alpha\beta} S_i^\alpha S_j^\beta,
 \label{H_eff}
\end{equation}
with anisotropic bilinear coupling $J_{ij}^{\alpha\beta}$, where $\alpha$ and $\beta$ label the three components of the pseudospin-$\sfrac{1}{2}$ ${\bm S}_i$ (see Appendix~\ref{sec:local})~\cite{ross2011quantum,savary2012order,rau2019frustrated,onoda2011quantumspinice,lee2012generic}. For non-Kramers systems, such as Tb$_2$Ge$_2$O$_7$, the $S_i^z$ component represents the time-odd magnetic moment operator while the transverse $S_i^\pm$ components track the electric quadrupole moment operator and other time-even multipoles~\cite{onoda2011quantumspinice,lee2012generic}. In Hamiltonian~\eqref{H_eff}, the dipolar $S_i^z$ and quadrupolar $S_i^\pm$ operators do not directly couple through a term of the form $S_i^z S_j^\pm$ since it would violate the time-reversal invariance of ${\cal H}_{\text{eff}}$~\cite{onoda2011quantumspinice,lee2012generic,rau2019frustrated}.

The central question is whether the physics of any of the terbium pyrochlores can be qualitatively captured by a model such as Eq.~(\ref{H_eff}) that neglects the excited crystal field doublet at $\Delta \sim 1.5$~meV, which is barely an order of magnitude larger than the energy scale of the ion-ion interactions given by the $J_{ij}^{\alpha\beta}$ couplings~\cite{takatsu2016quadrupole,kadowaki2015composite,gingras2000thermodynamic,kao2003understanding,molavian2007dynamically}. An immediate sign that Eq.~(\ref{H_eff}) is insufficient to describe Tb$_2$Ge$_2$O$_7$ comes from our magnetic structure determination, which revealed that the magnetic moments are \emph{strongly} canted away from the local $\langle 111 \rangle$ directions below $T^*=1.1$~K (see Sec.~\ref{sec:magstrucdeterm}). In the absence of excited crystal field states, Eq.~(\ref{H_eff}) requires that $g_\perp$ is strictly zero for non-Kramers ions~\cite{onoda2011quantumspinice,lee2012generic,rau2019frustrated}, meaning that the magnetic moments should point exactly along local $\langle 111\rangle$ directions in any dipole-ordered state that forms. 
Similar canting has also been observed in Tb$_2$Sn$_2$O$_7$, where it is argued to arise from the interaction-induced admixing between the crystal field ground and excited states~\cite{molavian2007dynamically,mcclarty2009energetic,mcclarty2010soft,petit2012spin2}. Such admixing, referred to as virtual crystal field fluctuations (VCFF)~\cite{rau2019frustrated,rau2016order}, introduces new terms in Hamiltonian~\eqref{H_eff}. Most importantly, VCFF lead to a coupling at the effective Hamiltonian level between dipolar and quadrupolar operators through three-spin interaction terms of the form $ \mathcal{K}_{ijk} (S_i^z S_j^\pm S_k^z + {\text {h.c.}})$. Below, we borrow the complementary mean-field theory and random phase approximation approach of Refs.~\cite{rau2016order,petit2012spin2,mcclarty2009energetic} to investigate the effects of interaction-induced admixing of crystal field ground and excited energy levels.

Considering the factors discussed above and the evidence assembled over the past twenty-five years in regards to the properties of the terbium pyrochlores, we reach the following conclusion: a minimal model that can provide a semi-quantitative description of Tb$_2$Ge$_2$O$_7$, and presumably all terbium pyrochlores, must contain three ingredients: 
(i) interactions between the angular momenta ${\mathbf J}_i$ that promote magnetic dipole order;
(ii) interactions between time-even multipoles (e.g electric quadrupoles) that
compete with the magnetic ordering; and 
(iii) a low-lying excited crystal field doublet of $\Delta = 1.5$ meV that allows interaction-induced admixing between the ground and excited doublet and intertwining of the magnetic dipoles and electric quadrupoles. Such a model is a necessary starting point to rationalize the whole terbium pyrochlore series and, thanks to the experimental insights reported in Section~\ref{sec:experiment}, we view Tb$_2$Ge$_2$O$_7$ as the linchpin for such an analysis.

\subsection{Model Hamiltonian}
\label{subsec:model-hamiltonian}

Our foremost goal is to expose theoretically the qualitative features (in terms of ordered phases and their inelastic neutron scattering response) that can arise from the competition between time-odd and time-even multipoles when the crystal field gap $\Delta$ is comparable in its energy scale to those interactions. For this purpose, we follow the spirit of Ref.~\cite{rau2016order}. We write a tripartite toy-model Hamiltonian that, in addition to the crystal field ${\cal H}_{\rm cf}({\mathbf J}_i)$, includes a term that serves as proxy for all interactions between time-odd multipoles, ${\cal H}_{\rm biEx}$ and another term for those between time-even multipoles, ${\cal H}_{\rm EQQ}$:
 \begin{equation}
{\cal H}=\sum_{i}{\cal H}_{\rm cf}({\mathbf J}_i)+{\cal H}_{\rm biEx}+\lambda\! \cdot \! {\cal H}_{\rm EQQ}.
\label{eq:Htot}
\end{equation}
Beyond the first excited crystal field doublet at $\Delta=1.5$~meV, the next highest excited crystal field levels are nearly an order of magnitude higher in energy (see Table~\ref{CEFenergy}). These higher energy states are thermally depopulated at the low temperatures where Tb-Tb correlations start to develop and can thus be ignored.
Moreover, in that way, we also disregard the interaction-induced admixing between the crystal field ground state and those high energy levels at energy $\gtrsim 10$ meV. Henceforth, we thus consider a reduced Hilbert subspace defined by the two lowest crystal field levels of Tb$_2$Ge$_2$O$_7$, whose wave functions are tabulated in Table~\ref{CEFcomp}.

In Eq.~\eqref{eq:Htot}, ${\cal H}_{\rm biEx}$ is a bilinear anisotropic exchange Hamiltonian expressed in terms of the components, ${\rm J}_i^u$, of the angular momenta, ${\mathbf J}_i$~\cite{ross2011quantum,curnoe2007quantum}~\footnote{Note that, here, the parameters $\cal J$'s (in calligraphic font) are bilinear ``exchanges'' between angular momentum operator ${\mathbf J}_i$. For simplicity, we ignore in this work the long-range part of the magnetic dipole-dipole interactions and incorporate its nearest-neighbor contributions into the ${\cal J}_{uv}$ bilinear couplings. As the nearest-neighbor ${\cal J}_{uv}$ can already generate a ferromagnetic $\Gamma_9$ long-range ordered canted spin ice phase such as observed in Tb$_2$Ge$_2$O$_7$, we leave the study of how dipolar interactions beyond nearest-neighbor affect the phase diagram that model (\ref{eq:Htot}) predicts for future studies.}, 
\begin{equation}
\begin{split}
&{\cal H}_{\rm biEx}=\sum_{\langle i, j \rangle}\Big\{{\cal J}_{zz}{\rm J}_i^z{\rm J}_j^z-{\cal J}_\pm ({\rm J}_i^+{\rm J}_j^-+\rm{h.c.})\\
&+{\cal J}_{\pm\pm}(\gamma_{ij}{\rm J}_i^+{\rm J}_j^+ + \rm{h.c.}) + 
{\cal J}_{z\pm}\left( \zeta_{ij}
[{\rm J}_i^z{\rm J}_j^+ + 
 {\rm J}_i^+{\rm J}_j^z] + \rm{h.c.} \right )
\Big\}.
\end{split}
\label{eq:HbiEx}
\end{equation}
Here ${\mathbf J}_i$ is the angular momentum operator expressed in its \textit{local} frame at site $i$, with $\hat{z}_i$ along the local $[111]$ axis with the sum running over nearest neighbors only. The complex phase factors $\gamma_{ij}$ are given in Appendix~\ref{sec:local} with $\zeta_{ij} = -\gamma_{ij}^*$.

To mimic the interactions between time-even multipoles and explore their effect on the thermodynamic phases
and dynamic response of Tb$_2$Ge$_2$O$_7$, we consider in
this work the electric quadrupole-quadrupole interaction, ${\cal H}_{\rm EQQ}$~\cite{Finkelstein1953interatomic,bleaney1961quadrupole,morin1989interplay,rau2016order},
\begin{equation}
\begin{split}
&{\cal H}_{\rm EQQ}=\\
&{\mathscr Q}\sum_{\langle i,j\rangle}\hat{\bm r}_{ij}^T\left(\frac{1}{6}{\rm Tr}[\bm Q_i\bm Q_j]-\frac{10}{6}\bm Q_i\bm Q_j+\frac{35}{12}\bm Q_i\hat{\bm r}_{ij}\hat{\bm r}_{ij}^T\bm Q_j\right)\hat{\bm r}_{ij} .
\end{split}
\label{eq:HEQQ}
\end{equation}
Here, ${\bm Q_i}$ is the quadrupole moment operator, which is a rank-2 tensor, defined with respect to the \textit{global} frame as $Q_{i}^{\alpha\beta}
\equiv\frac{3}{2}({\rm J}^\alpha_i {{\rm J}}^\beta_i + { {\rm J}}^\beta_i {{\rm J}}^\alpha_j)
-{{\rm J}}({{\rm J}}+1)\delta_{\alpha\beta},$ with angular moment ${\mathbf J}_i$ in the \textit{global} frame at the pyrochlore lattice site ${\bm r}_i$, with 
${\bm r}_{ij}\equiv {\bm r}_i-{\bm r}_j$, and $\alpha,\beta=x,y,z$. For Tb$^{3+}$, which has ${{\rm J}}=6$, the EQQ coupling constant is ${\mathscr Q}\simeq 1.635\times10^{-3}$ K
\footnote{The EQQ coupling constant~\cite{wolf1968electric} ${\mathscr Q}=e^2\alpha^2{\langle {r^2}\rangle}^2/(4\pi\epsilon_0 r_{\rm nn}^5)$, with ${\langle {r^2}\rangle}$ is the mean-square $f$-electron radius of Tb$^{3+}$~\cite{freeman1979dirac-fock}, $\alpha$ is the reduced matrix element as defined in Ref.~\cite{stevens1952matrix}, $r_{\rm nn}$ is the separation between nearest neighbor Tb$^{3+}$, which we use 3.535 \AA, and $\epsilon_0$ is the permittivity of free space.}. As the EQQ interaction decays rapidly as $1/{r_{ij}}^5$, we also consider only the nearest-neighbor contribution of Eq.~\eqref{eq:HEQQ}. In Eq.~\eqref{eq:Htot}, $\lambda\in[0,1]$ is a dimensionless scale, which controls the screening of the Coulomb interaction at the 
origin of ${\cal H}_{\rm EQQ}$~\footnote{Screening from the outer closed shells ($5s^25p^6$) normally reduces the quadrupole moment~\cite{wolf1968electric} while coupling with the open-shell electrons, especially the $5d$, increases it~\cite{levy1979large}. 
Note also that Eq.(~\ref{eq:HEQQ}) differs by an overall factor $\sfrac{1}{2}$ compared to
the corresponding ${\cal H}_{\rm EQQ}$ in Ref.~\cite{rau2016order} where this factor was accidentally
omitted.}.

Similarly to the bilinear interactions ${\cal H}_{\rm biEx}$ \eqref{eq:HbiEx}, it is convenient to work in the local orthogonal $(x_i,y_i,z_i)$ frame at site $i$ and rewrite ${\cal H}_{\rm EQQ}$ in Eq.~\eqref{eq:HEQQ} in terms of Stevens operators, $O_2^\mu(i)$~\cite{jensen1991}, expressed as
\begin{equation}
{\cal H}_{\rm EQQ}=\sum_{\langle i, j \rangle}\sum_{\mu,\nu}{\cal M}_{ij}^{\mu\nu}O_2^\mu(i)O_2^\nu(j) .
\label{eq:Hq}
\end{equation}
Here, $\mu,\nu=2,1,0,-1,-2$, index the five rank-2 (quadrupole) Stevens operators $O_2^\mu(i)$.
See Appendix~\ref{sec:Meqq} for the expression of $O_2^\mu(i)$ in terms of the local frame components of ${\mathbf J}_i$ and for details of the interaction matrix ${\cal M}_{ij}^{\mu\nu}$.

\begin{figure*}[tbp]
\linespread{1}
\par
\includegraphics[width=7in]{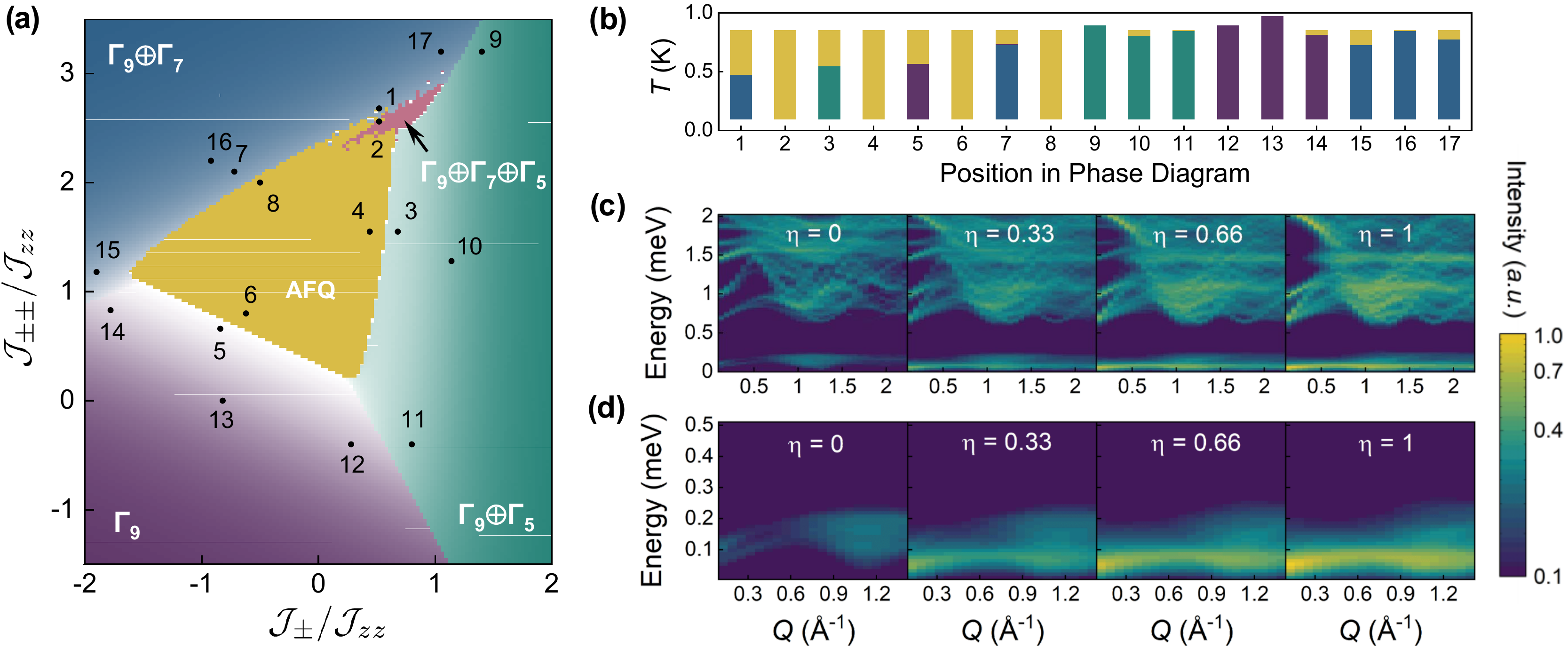}
\par
	\caption{ (Color) 
	(a) Mean-field ground state phase diagram of a two-doublet non-Kramers system on the pyrochlore lattice with dipolar and quadrupolar interactions given by Eq.~\eqref{eq:Htot}. We have selected parameters that are relevant to Tb$_2$Ge$_2$O$_7$: $\Delta=1.5$~meV, ${\cal J}_{zz}=0.012$~meV, ${\cal J}_{z\pm}=2/3{\cal J}_{zz }$, and $\lambda=0.28$ (see Appendix~\ref{sec:fit}). The brightness of the white band surrounding the AFQ phase is proportional to the relative weight of the $\Gamma_{9-\rm SI}$ configuration where a higher degree of whiteness signifies a smaller canting angle away from the local $\langle 111\rangle$ directions. There are five distinct ordered phases and each dipole phase is associated with a quadrupole order parameter, as described in Table~\ref{tab:di-qua}. 
	(b) Finite-temperature phase diagram for locations points 1 to 17 in (a), with the same color coding. The system is in a paramagnetic phase above $T\gtrsim 0.8$ K ($\pm 0.1$ K, depending on the parameters). At some locations in the phase diagram there is an intermediate long-range AFQ region, represented by a yellow band, which is the same phase as the central region of panel (a).  
	(c,d) Spin wave calculations with RPA showing that our inelastic neutron scattering data for Tb$_2$Ge$_2$O$_7$, at temperatures $T_c < T < T_\Delta$, is qualitatively consistent with a coexistence of dipolar and quadrupolar inelastic neutron scattering signal. We superimpose the inelastic neutron scattering spectra at points 1 and 2 in (a), which are in the dipolar ($I_{\rm MD}(Q,\omega)$) and quadrupolar ($I_{\rm EQ}(Q,\omega)$) phases, respectively: $\tilde I(Q,\omega;\eta)=(1-\eta) I_{\rm MD}(Q,\omega) +\eta I_{\rm EQ}(Q,\omega)$ where $\eta$ is the weight of the quadrupolar correlations.} 
	\label{fig:pd+ins}
\end{figure*}

In the two sections that follow, we investigate candidate ground states and phase diagram that result from model ~\eqref{eq:Htot} along with the associated spin dynamics using mean-field and random-phase approximations (MF-RPA)~\cite{jensen1991,rau2016order,petit2012spin}. We implement a mean-field procedure in terms of the expectation values of local magnetic dipole (MD) and electric quadrupole (EQ) moments, $\langle {\mathbf J}_i\rangle$ and $\langle O_2^\mu(i) \rangle$, respectively. We search for ${\mathbf {k}}={\mathbf 0}$ orders by solving the self-consistent MF equations (see Appendix~\ref{sec:MF-RPA}) for $\langle {\mathbf J}_i\rangle$ and $\langle O_2^\mu(i)\rangle$ within a single tetrahedron~\footnote{This is because we have a 4-site (tetrahedron) basis for ${\mathbf{k}}=
 {{\mathbf 0}}$ ordering. This should not be confused with a cluster mean-field calculation such as done in Ref.~\cite{javanparastObD2015} for a model of Er$_2$Ti$_2$O$_7$, for example.}, choosing the solution that minimizes the free energy in Eq.~\eqref{eq:free-energy}. See Appendix~\ref{sec:order-params} for a definition of the order parameters characterizing the dipolar and quadrupolar phases of model ~\eqref{eq:Htot}. We then discuss the findings of our MF-RPA calculations in relation to our experimental observations of Tb$_2$Ge$_2$O$_7$.~\footnote{The derivation of a pseudospin-$\sfrac{1}{2}$ from model ~(\ref{eq:Htot}) using perturbation theory will be presented elsewhere~\cite{wonglost2020}.}.

\subsection{Mean-field phase diagram}
\label{subsec:mft-phase-diagram}

We begin by discussing ground state phases of model \eqref{eq:Htot}. A typical ground state phase diagram arising from the competition between the interactions defining Eq.~(\ref{eq:Htot}) is illustrated in Fig.~\ref{fig:pd+ins}(a) (see Appendix~\ref{sec:fit} for a discussion of the choice made for the ${\cal J}_{uv}$ parameters of Eq.~(\ref{eq:HbiEx})). Despite being a simplification of the real Tb$_2$Ge$_2$O$_7$ Hamiltonian, this model already displays a much increased richness of competing ground state phases compared to the heretofore considered pseudospin-$\sfrac{1}{2}$ model of non-Kramers ions~\cite{rau2019frustrated,lee2012generic,onoda2011quantumspinice,kadowaki2015composite,takatsu2016quadrupole}. The phase diagram contains five distinct ordered phases as defined in Table~\ref{tab:di-qua}.

The first region we discuss, indicated in yellow in Fig.~\ref{fig:pd+ins}(a), is a phase with long-range quadrupolar order in which the magnetic dipoles remain disordered. The quadrupolar ordered phases that we encounter in this work are described as either ferroquadrupolar (FQ) or antiferroquadruplar (AFQ). The FQ phases have the same orientation of the principal axes of the quadrupoles on all four sublattices, as the name implies. The AFQ phases are described by having pairs of quadrupole principal axis orientations at two of four sublattices identical, with the other pair on a tetrahedron being different. The quadrupolar phase in the central yellow region of Fig.~\ref{fig:pd+ins}(a) is antiferroquadruplar. In this region of the phase diagram, the molecular mean-field, which is entirely of EQQ origin, splits the single-ion crystal field ground state doublet and entangles the \ket{1{\rm st}} and \ket{2{\rm nd}} crystal field states (see Table~\ref{CEFcomp}). The mutually frustrated bilinear ${\cal J}_{uv}$ couplings are unable to overcome the energy gap generated by the collective quadrupolar ordering and therefore remain disordered. Note that this AFQ phase without dipolar order is absent at $\lambda=0$ for the range of ${\cal J}_{uv}$ values 
considered~\footnote{Consider the perspective of an effective pseudospin-$\sfrac{1}{2}$ Hamiltonian, such as \eqref{H_eff},
derived using perturbation theory~\cite{molavian2007dynamically,rau2016order}. In such a case, for the model~\eqref{eq:Htot} with ${\cal H}_{\rm biEx} \ne 0$ and $\lambda=0$, in addition to the approximation of infinite gap $\Delta \rightarrow \infty$, an effective pseudospin-$\sfrac{1}{2}$ Ising model with $J_{zz}\ne 0$, $J_{\pm}= 0$ and $J_{\pm\pm}= 0$ in Eq.~\eqref{H_eff} would be realized. Nonzero $J_{\pm}$ and $J_{\pm\pm}$ couplings would be reintroduced only perturbatively, to the order of $1/\Delta$. Considering classical ground states only, for the choice of ${\cal J}_{uv}$ parameters made (in Eq.~\eqref{eq:HbiEx}), such VCFF-induced $J_\pm$ and $J_{\pm\pm}$ remain perturbatively small compared to $J_{zz}$ and dipolar $J_{zz}$-induced correlations ensues. 
Predominant quadrupolar order only wins in model ~\eqref{eq:Htot} for a choice of ${\cal J}_{uv}$ if $\lambda$ is greater than a critical value $\lambda_c$.}. This regime of strong quadrupolar correlations may be relevant to off-stoichiometric Tb$_{2+x}$Ti$_{2-x}$O$_{7+y}$, which has been claimed to realize an electric quadrupole phase~\cite{lee2012generic,kadowaki2015composite,takatsu2016quadrupole}. This lends credence to the hope that our model~\eqref{eq:Htot} can expose key aspects of the generic physics at play in Tb$_2$Ge$_2$O$_7$ and Tb pyrochlores in general.

Four dipole ordered ground state phases surround the AFQ phase in the phase diagram of Fig.~\ref{fig:pd+ins}(a), each with an associated quadrupolar ordering described in Table~\ref{tab:di-qua}. In the primary dipolar ordered phases, it is somewhat redundant to expand much on the accompanied and enslaved quadrupole orders as they develop simultaneously with the primary dipolar orderings. That being said, it is important to include the quadrupolar molecular field 
when describing the spin dynamics within the dipole ordered phases (see Appendix~\ref{app:spin_dynamics}). These four dipolar phases appear when the anisotropic bilinear exchanges ${\cal J}_{u\hspace{0.5pt}v}$ are sufficiently large compared to the energy scale of the quadrupolar interactions, $\lambda {\mathscr Q}$. For all these phases, we observe a nonzero transverse ($xy$) component to the magnetic moment, meaning that the ordered dipole moment is canted away from its local $[111]$ direction. A nonzero transverse magnetic moment is also preserved when $\lambda=0$~\cite{mcclarty2010soft}. 
Thus, interaction-induced admixing between two doublets~\cite{rau2019frustrated,rau2016order,molavian2007dynamically,mcclarty2009energetic,mcclarty2010soft,petit2012spin2} plays an important role in generating a transverse ($xy$) dipole moment for non-Kramers ions with zero transverse single-ion anisotropy, $g_\perp=0$~\cite{molavian2007dynamically}.

In order to describe the four dipole ordered phases, we decompose the total magnetic moment in terms of its local $z$ and $xy$ components. Depending on the sign of the ${\cal J}_{zz}$ coupling, and up to order $(1/\Delta)$, there are two possible configurations for the $z$ components: all-in/all-out order (AIAO, $\Gamma_3$) or two-in/two-out ordered spin ice ($\Gamma_{9-\rm SI}$) ~\cite{den2000dipolar}. For example, for the phase labelled $\Gamma_9\oplus \Gamma_7$, (henceforth, by $\Gamma_9$, we mean both $\Gamma_{9-\rm SI}$ combined with $\Gamma_{9-\rm SFM}$, see Table ~\ref{tab:orders}), the $xy$ components of the magnetic moment are made up of both the Palmer-Chalker $\Gamma_7$ and splayed ferromagnet (SFM) $\Gamma_{9-\rm SFM}$ components. To characterize the extent of the canting of the dipole moment away from the local $\langle$111$\rangle$ directions, we compute the ``weight'' of the SI ($\Gamma_{9-\rm SI}$) component in the magnetic moment configuration~\footnote{We define this weight as $w_{\rm SI}\equiv\frac{1}{A}\cdot\bm m_{\Gamma_{9-\rm SI}}^2$, with $A\equiv (\bm m_{\Gamma_3}^2+\bm m_{\Gamma_5}^2+\bm m_{\Gamma_7}^2+\bm m_{\Gamma_{9-\rm SFM}}^2+\bm m_{\Gamma_{9-\rm SI}}^2)$}, 
which is illustrated in Fig.~\ref{fig:pd+ins}(a) by the degree of ``whiteness'' in the dipole ordered phases and is largest on the boundaries with the central AFQ phase.

\begin{table}[t]
	\caption{\label{tab:di-qua} Description of the five ordered ground state phases in Fig.~\ref{fig:pd+ins}(a). There is one purely quadrupolar ordered phase lacking dipolar order and four dipolar phases. Each dipolar phase is associated with a particular type of quadrupolar order parameter. The abbreviation FQ stands for ferroquadrupolar and AFQ for antiferroquadrupolar. In the $\Gamma_9\oplus\Gamma_5$ and $\Gamma_9\oplus\Gamma_7\oplus\Gamma_5$ phases, the orientation of the principal axes of the quadrupoles is different on each of the four sublattices and the quadrupole ordering therefore has
	mixed AFQ and FQ character, hence the label AFQ $\oplus$ FQ. }
	\begin{tabular}{c|c|cccc}
		\hline\hline
		&\begin{tabular}{c}
			Quadrupolar \\phase
		\end{tabular}&\multicolumn{4}{c}{Dipolar phase}\\
		\hline
		\begin{tabular}{c}
		 Dipole \\ ordering
		\end{tabular}& Disordered&$\Gamma_9$&$\Gamma_9\oplus\Gamma_7$&$\Gamma_9\oplus\Gamma_5$&$\Gamma_9\oplus\Gamma_7\oplus\Gamma_5$\\
		\hline
		\begin{tabular}{c}
			Quadrupole \\ordering$^a$
		\end{tabular}
		&AFQ&FQ&FQ&AFQ $\oplus$ FQ&AFQ $\oplus$ FQ\\
		\hline\hline
	\end{tabular}
\vspace{1ex}

 {\raggedright $^a$ We emphasize that this is a description of the orientation of the quadrupoles in the four dipolar ordered ground states of Fig.~\ref{fig:pd+ins}(a). At many locations in the phase diagram, the dipole order is preceded by long-range AFQ order (yellow bands in Fig.~\ref{fig:pd+ins}(b)). \par}
\end{table}

A particularly interesting region of the phase diagram, indicated by the sliver of pink in the top right of Fig.~\ref{fig:pd+ins}(a), is a small region in which the dipole order parameter is described by the combination of the three $\Gamma_9\oplus\Gamma_7\oplus\Gamma_5$ components. Due to the superposition of both $\Gamma_5$ and $\Gamma_7$, the magnitude of the local transverse ($xy$) moment differs on the four sublattices and the net ordered magnetic moments in this phase are therefore inequivalent on each of the four sublattices. The development of this $\Gamma_9\oplus\Gamma_7\oplus\Gamma_5$ phase with
unequal moment on each of the four sublattices may suggest that the system would rather develop ${\mathbf{k} }\neq {\mathbf 0}$ ordering~\cite{yan2017theory} than the ${\mathbf {k}}={\mathbf 0}$ solution sought here. This could be relevant to the $\mathbf{k}=(0,0,1)$ phase that emerges in Tb$_2$Ge$_2$O$_7$ below $T_c=0.25$~K.

The ground state phase diagram of Fig.~\ref{fig:pd+ins}, already richer in the number and the complexity of phases it displays compared to that of the well-studied pseudospin-$\sfrac{1}{2}$ model~\cite{rau2019frustrated,onoda2011quantumspinice,lee2012generic}, is also more complex in its finite temperature behavior. The finite-temperature phase diagram of the model given by Eq.~\eqref{H_eff} displays very limited regions in $J_{ij}^{\alpha\beta}$ parameter space where temperature-driven phase transitions occur \emph{between} long-range ordered phases ~\cite{jaubert2015multiphase,yan2017theory,taillefumiercompeting2017}. In contrast, the phase diagram in Fig.~\ref{fig:pd+ins}(a) harbors more intricate phase behavior~\cite{fernandes2019intertwined} at finite temperature (Fig.~\ref{fig:pd+ins}(b)). We first consider the temperature-dependent phases displayed by systems at points 1 to 8 in Fig.~\ref{fig:pd+ins}(a), which are located near the boundary between dipole ordered and dipole disordered (AFQ) phases. For points 1, 3, 5 and 7, all of which have dipole ordered ground states, we find that the system enters into its dipole ordered phase via a two step process that involves an intermediate AFQ phase over an extended temperature window (see Fig.~\ref{fig:pd+ins}(b)). For systems located at points 2, 4, 6 and 8, only AFQ order occurs (recalling that quantum fluctuations, which have the potential to destroy long-range ordered phases~\cite{onoda2011quantumspinice,lee2012generic} and give rise to a $U(1)$ quantum spin liquid~\cite{gingras2014quantum}, are not included here). For systems that sit far from the AFQ phase (\emph{e.g.} points 9--13), deep inside dipole ordered phases, an intermediate AFQ phase is either absent (\emph{e.g.} points 12 and 13) or limited temperature extent (\emph{e.g.} points 10 and 11). For all 17 locations marked in Fig.~\ref{fig:pd+ins}(a), the system is paramagnetic above $T \gtrsim 0.8 \pm 0.1$ K (see Fig.~\ref{fig:pd+ins}(b)).

\subsection{Inelastic neutron scattering -- overall perspective}
\label{subsec:INS-perspective}

Having found many states that naturally arise and compete in a toy model pertinent to Tb$_2$Ge$_2$O$_7$, we next turn to the question of the theoretical inelastic neutron scattering for the phases of Fig.~\ref{fig:pd+ins}(a) in relation to the experimental data of Fig.~\ref{DCS_Slices}. We first recap the essential experimental results. Below $T_\Delta = 6$~K, inelastic scattering intensity centered at energy $0.18$ meV and momentum $1.2$ {\AA}$^{-1}$ begins to grow (green diamonds in Fig.~\ref{fig:Summary}). Because the magnetic dipolar transition matrix elements between the two states (\ket{1{\rm st}} and \ket{2{\rm nd}}, see Table~\ref{CEFcomp}) of the crystal field ground state doublet vanish, we associate the visibility of this inelastic signal with extended quadrupolar correlations that entangle the two states of the ground state doublet~\cite{petit2012spin2}. These correlations, as indicated by the inelastic signal, continue to grow until $T^*=1.1$~K without an obvious thermodynamic feature ({\emph{e.g.}} in heat capacity) that would signal spontaneous long-range quadrupolar order, as occurs in a number of the scenarios illustrated in Fig.~\ref{fig:pd+ins}(b). Below $T^*$, the growth of the inelastic signal slows down and eventually saturates at $0.6$~K. Finally, at $T_c=0.25$~K, a genuine thermodynamic transition occurs, accompanied by a drop in the inelastic signal at $1.2$ {\AA}$^{-1}$.

We thus arrive at the key question: what is the nature of the state in the regime 
$T_c \! < \! T \! < \! T_\Delta$? The magnetic (dipolar) character of this correlated state reminds one of the spin ice regime in Ising spin ice models with either long range dipolar interactions~\cite{melko2001DSILRO,melko2004DSIMC} or weak exchange beyond nearest neighbors~\cite{applegate2012vindication,rau2016spin}. However, in dipolar spin ices, such as Ho$_2$Ti$_2$O$_7$ and Dy$_2$Ti$_2$O$_7$, the correlated spin ice regime is essentially invisible in the inelastic channel~\cite{clancy2009revisiting}. This is due to the the non-Kramers nature of Ho$^{3+}$ in the former, the highly protected Ising character of the crystal field ground doublet~\cite{rau2015quantumeffects,clancy2009revisiting,Gaudet2018vibron} in the latter and the large crystal field gap $\Delta \sim 30$ meV in both compounds, all making the aforementioned magnetic dipole transition matrix elements extremely small. Here, for Tb$_2$Ge$_2$O$_7$, the low-lying excited crystal field level at $\Delta = 1.5$~meV allows the dipolar and quadrupolar correlations to strongly intertwine. We thus argue that this temperature regime consists of a strongly correlated collective paramagnet where dipolar and quadrupolar correlations each contribute their \emph{respective} signatures to the neutron scattering, $I(Q,\omega)$. With this perspective in place, we next discuss how such a scenario may be qualitatively captured theoretically.

\subsection{Inelastic neutron scattering -- modeling}
\label{subsec:INS-theory}

A theoretical computation of the inelastic neutron scattering intensity in the correlated liquid state of Tb$_2$Ge$_2$O$_7$ ($T_c < T < T_\Delta$) would be an extremely difficult task. Furthermore, even with the minimal nature of the model considered
(Eq.~(\ref{eq:Htot})), a quantitative determination of the parameters
$\{
{\cal J}_{zz}, {\cal J}_{\pm}, {\cal J}_{z\pm}, 
{\cal J}_{\pm\pm},\lambda
\}$
would require experiments on single crystal samples, which are not currently available. Nevertheless, we wish to illustrate that the above physical picture has some theoretical underpinning. In this section, we present random phase approximation (RPA) calculations of the powder-averaged $I(Q,\omega)$ within a scheme that qualitatively describes a state with co-evolving dipolar and quadrupolar correlations.

In order to narrow down the regions of the phase diagram in Fig.~\ref{fig:pd+ins}(a) physically relevant to Tb$_2$Ge$_2$O$_7$, we consider two experimental observations: the splitting of the ground state doublet, 
$E_{\text{gap}}=0.18\pm 0.015$~meV, and the canting angle of the dipole moment away from the local $1 111]$, $24.5 \pm 1.5^{\circ}$. Using these two values, we apply a two-parameter constraint (for the set of parameters chosen to produce the phase diagram of Fig.~\ref{fig:pd+ins}(a)) and find that both are satisfied within the dashed white (arrowhead shaped) wedges of Fig.~\ref{fig:fit} in Appendix~\ref{sec:fit}, giving ${\cal J}_{z\pm} \approx 2{\cal J}_{zz}/3$ and $\lambda\approx 0.28$. We refer the reader to Appendix ~\ref{sec:fit} for a detailed discussion motivating the choice of the location in the phase diagram we select for the RPA computation of $I(Q,\omega)$ and comparison with the experimental results shown in Fig.~\ref{DCS_Slices}.

We next compute $I(Q,\omega)$ at the candidate location point 1 in Fig.~\ref{fig:pd+ins}(a), where $({\cal J}_{\pm},{\cal J}_{\pm\pm})=(0.00624, 0.03072)$ meV, which is in the $\Gamma_9\oplus\Gamma_7$ dipole ordered phase. This is shown in the leftmost, $\eta=0$, panels of Figs.~\ref{fig:pd+ins}(c,d) for two different maximum energy transfers to facilitate comparison with Fig.~\ref{DCS_Slices}. The broad excitation spanning the range 0.5 to 2~meV corresponds to the first excited crystal field level, which picks up significant dispersion due to the
$({\cal H}_{\rm biEx}+{\cal H}_{\rm EQQ})$ interactions~\cite{kao2003understanding}. 
Consistent with the experimental data, the computed $I(Q,\omega)$ at this location displays a broad intensity maximum around $Q = 1.2~{\AA}^{-1}$ and is roughly centered at $E = 0.18$ meV. The intensity drops in the region near $0.6 ~{\AA}^{-1}$ and re-intensifies below $Q = 0.3~{\AA}^{-1}$. In this computed spectrum, the overall intensity of the collective mode centered at 0.18~meV is at most on par with the intensity of the crystal field level near 1.5~meV whereas in the experimental data the intensity of the collective excitation dwarfs that of the crystal field (Fig.~\ref{DCS_Slices}(e)).

It is important to emphasize that the inelastic mode at 0.18~meV would be essentially invisible in the limit of a well-isolated crystal field ground state ($\Delta \rightarrow \infty$), because, in such a scenario, 
dipolar and quadrupolar order parameters do not co-exist~\footnote{Note though that even in a case with $\Delta \rightarrow \infty$, one might expect thermal and quantum fluctuation to generate nonlinear interactions
(\emph{e.g.} in a Ginzburg-Landau description) between dipolar and quadrupolar order parameters, tough these should be in some sense small for lack of their coupling at the Hamiltonian level when $\Delta = \infty$.}.
 The finite $\Delta$ present here induces a secondary quadrupolar order parameter in $\Gamma_9\oplus\Gamma_7$ (see Table ~\ref{tab:di-qua}) producing a partial admixing of the (otherwise strictly Ising) crystal field ground states (\ket{1{\rm st}} and \ket{2{\rm nd}}) along with an admixing between the ground and excited doublets~\cite{petit2012spin2}. Both effects contribute to the visibility of the mode at 0.18~meV. In short, strong and predominant dipolar order on its own (see Fig.~\ref{fig:pd+ins}(c,d), $\eta=0$ panel) produces an inelastic response that is weakly visible for a non-Kramers system. The same qualitative behavior is found in all locations of the two dashed white bands of Fig.~\ref{fig:fit} in Appendix ~\ref{sec:fit}.

To model a regime of co-evolving intertwined dipolar and quadrupolar correlations, as occurs in the $T_c < T < T_\Delta$ temperature interval, we next consider point 2, which is in close vicinity to point 1 considered above, but now, just beyond the boundary and barely within the AFQ phase (central
yellow phase of Fig.~\ref{fig:pd+ins}(a)). This is illustrated in panels (c) and (d) of Fig.~\ref{fig:pd+ins} for $\eta=1$. We immediately see the impact of quadrupolar correlations in generating intensity at low energies and $Q = 1.2 \AA^{-1}$ through their ability to admix the \ket{1{\rm st}} and \ket{2{\rm nd}} states. However, as point 2 is barely in the AFQ phase, the quadrupolar-induced splitting for $\eta=1$ is small and shifted downwards in energy to a value of $\sim 0.05$~meV while the entanglement of \ket{1{\rm st}} and \ket{2{\rm nd}} is maximal. We thus see that by combining the scattering arising from both strong dipolar and quadrupolar order ($\eta=0.33$ and $\eta =0.66$), so as to mimic their coexistence, we can qualitatively capture ($\eta=0.66$) inelastic scattering that bears intriguing similarities to the experimental results in Fig.~\ref{DCS_Slices} in the temperature interval 0.45 K $\le T \le $ 3 K. 
Since our computed $I(Q,\omega;\eta)$ assumes either a long-range dipolar ($\eta=0$) or quadrupolar ($\eta=1$) ordered phase, it is not surprising that features in the spectral weight are narrower in energy than what is found experimentally in the $T_c<T<T_\Delta$ collective paramagnetic regime. However, it is surprising to note that the experimental inelastic signal remains very broad in energy even in the ordered phase ($T < T_c=0.25$~K, see Fig~\ref{DCS_Slices}(j-l)), which could perhaps be explained by the size of the magnetically ordered domains.

Our MFT+RPA calculations show convincingly that the complex phase evolution in Tb$_2$Ge$_2$O$_7$ arises naturally from the opportune convergence of two key aspects of the Tb$^{3+}$ ions in \tgo{}: (i) strong dipolar and quadrupolar interactions that mutually compete and promote their respective and distinct spatial and dynamical correlations and (ii) a low-lying excited crystal field doublet of $\Delta = 1.5$ meV that provides a channel for interaction-induced admixing between the ground and excited doublets that intertwines the dipolar and quadrupolar interactions and the correlations they drive. While our calculation cannot yet provide a fully quantitative description of the inelastic scattering in Tb$_2$Ge$_2$O$_7$, and other terbium pyrochlores, it nonetheless represents a major step forward in our understanding of this
 fascinating family of compounds.



\section{Open Questions and Conclusions}
\label{sec:discussion}

In light of the theoretical results obtained in the previous section, we return to the interpretation of our experimental results. We observed three distinct magnetic phases in the pyrochlore Tb$_2$Ge$_2$O$_7$ that can be identified via three thermodynamic anomalies in its heat capacity. The first thermodynamic anomaly, at $T_{\Delta}=6$~K, originates from the thermal depopulation of a low-lying crystal field level, and also corresponds to the temperature scale on which collective spin excitations begin to form. The intensity of this collective mode can only be explained by the onset of quadrupolar correlations that enhances the visibility of this inelastic scattering. In the intermediate temperature regime, below $T^* =1.1$~K, we observed the formation of $\mathbf{k}={\mathbf {0}}$ quasi-Bragg peaks due to partial ordering of the dipole moments into a canted spin ice state. In response to the formation of dipolar order, the growth of \emph{independent} quadrupolar correlations is arrested. Below $T_c =0.25$~K, we detect new Bragg peaks at $\mathbf{k}=(0,0,1)$ positions that also correlate with an increase in the strength of the $\mathbf{k}={\mathbf {0}}$ quasi-Bragg peaks, while decreasing the strength of the quadrupolar correlations as reflected by the intensity of the low energy inelastic scattering. 

Significantly, only the lowest temperature of these three thermodynamic anomalies, at $T_c =0.25$~K, is sufficiently sharp in temperature to signify a true phase transition. In and of itself, this raises several unanswered questions regarding the nature of Tb$_2$Ge$_2$O$_7$ for $T_c$ 
$\le T \le T^*$. Our minimal model Hamiltonian naturally leads to competing quadrupolar and dipolar ordered phases, with the dipolar ordered phase winning out below $T_c$. But why is a quadrupolar long-range ordered phase not observed in the intermediate regime? Why is the thermodynamic anomaly at $T^*$ not a conventional phase transition, and why does Tb$_2$Ge$_2$O$_7$ display only short range elastic correlations for $T_c \le T \le T^*$?

Our experimental and theoretical work strongly supports the existence of phase competition between dipolar and quadrupolar order in Tb$_2$Ge$_2$O$_7$. It also provides a natural vehicle to explain the complexity of the phase behavior in terbium pyrochlores. The strong stoichiometry dependence observed in Tb$_2$Ti$_2$O$_7$ can be understood as originating from this compound lying close to the boundary of two adjacent phases, which given the richness of the phase diagram uncovered in this work, is far from unlikely. Furthermore, some samples of Tb$_2$Ti$_2$O$_7$, those that display no dipole ordering, appear to be consistent with an AFQ phase. Tb$_2$Sn$_2$O$_7$, which has the highest ordering transition is likely furthest from a dipolar/quadrupolar phase boundary. While the terbium pyrochlores and particularly Tb$_2$Ti$_2$O$_7$ have received decades of experimental investigation, progress has been slower on the theoretical front. We view Tb$_2$Ge$_2$O$_7$, a material where the intertwined nature of the competing orders is so clearly expressed as three distinct low temperature regimes, as the linchpin in achieving a comprehensive understanding of the terbium pyrochlores and for which we have laid the groundwork here.


\begin{acknowledgments}

We thank Paul McClarty, Jeffrey Rau and Daniel Wong for useful and stimulating discussions. This work was supported by the Natural Sciences and Engineering Research Council of Canada (NSERC). A.M.H. acknowledges support from the Vanier Canada Graduate Scholarship Program and thanks the National Institute for Materials Science (NIMS) for their hospitality and support through the NIMS Internship Program. M.J.P.G. and C.R.W. acknowledge support through the Canada Research Chairs Program (Tier I and Tier II, respectively). C.R.W. acknowledges support from the Leverhulme Trust. Work at the NIST Center for Neutron Research is supported in part by the National Science Foundation under Agreement No. DMR-0944772. This research used resources at the Spallation Neutron Source, a DOE Office of Science User Facility operated by the Oak Ridge National Laboratory (ORNL). We acknowledge the support of the National Institute of Standards and Technology, U.S. Department of Commerce, in providing the neutron research facilities used in this work.

\end{acknowledgments}

\bibliography{TbGeO_Ref_2017}

\begin{thebibliography}{108}
\bibitem [{CFm()}]{CFmoment}%
  \BibitemOpen
  \href@noop {} {}\bibinfo {note} {We note that while the strictly Ising nature
  and dominant \ket{{\text{J}}^z=\pm4} composition of the CF ground state are
  robust conclusions of our CF analysis, the exact magnitude of the CF ground
  state moment has a relatively large uncertainty. This uncertainty arises
  because its magnitude is very sensitive to the exact weighting of the
  \ket{{\text{J}}^z=\pm4} and \ket{{\text{J}}^z=\mp5} eigenstates in the CF
  ground state, which in turn are sensitive to the exact CF parameters that we
  refined. For a reasonable estimate of the uncertainty, we have calculated the
  CF moment obtained for an ensemble of CF parameters that has a calculated
  $\chi^2$ that can be 1/N$_{{\rm obs}}$ higher than the best solution, where
  N$_{{\rm obs}} = 10$ is the number of observables used to constrain the fit.
  This procedure leads to an estimated CF ground state moment of
  2.9(9)$\mu_B$.}\BibitemShut {Stop}%
\end{thebibliography}

\clearpage

\appendix

\section{Local frame}
\label{sec:local}

In this paper, we follow the conventions of Refs.~\cite{savary2012order,rau2016order} and work with a local frame of orthonormal axes for each of the four sites of the tetrahedral basis. Expressed in the global frame, the local axes for the four sublattices are defined in terms of their $(\hat x$, $\hat y$ and $\hat z$ components as
\begin{equation}
\begin{split}
\vhat{z}_1 &= \frac{1}{\sqrt{3}} \left(+\vhat{x}+\vhat{y}+\vhat{z}\right), \quad
\vhat{x}_1 = \frac{1}{\sqrt{6}} \left(-2\vhat{x}+\vhat{y}+\vhat{z}\right), \\
\vhat{z}_2 &= \frac{1}{\sqrt{3}} \left(+\vhat{x}-\vhat{y}-\vhat{z}\right), \quad
\vhat{x}_2 = \frac{1}{\sqrt{6}} \left(-2\vhat{x}-\vhat{y}-\vhat{z}\right), \\
\vhat{z}_3 &= \frac{1}{\sqrt{3}} \left(-\vhat{x}+\vhat{y}-\vhat{z}\right), \quad
\vhat{x}_3 = \frac{1}{\sqrt{6}} \left(+2\vhat{x}+\vhat{y}-\vhat{z}\right), \\
\vhat{z}_4 &= \frac{1}{\sqrt{3}} \left(-\vhat{x}-\vhat{y}+\vhat{z}\right), \quad
\vhat{x}_4 = \frac{1}{\sqrt{6}} \left(+2\vhat{x}-\vhat{y}+\vhat{z}\right), 
\end{split}
\label{eq:global-local}
\end{equation}
and $\vhat{y}_i = \vhat{z}_i \times \vhat{x}_i$. The bond phase
factors $\gamma_{ij}$ in ${\cal H}_{\rm biEx}$~\eqref{eq:HbiEx} is 
\begin{equation}
\label{eq:gamma_eq}
\gamma = \left(\begin{tabular}{cccc}
$0$ & $+1$ & $\omega$ & $\omega^2$ \\
$+1$ & $0$ & $\omega^2$ & $\omega$ \\
$\omega$ & $\omega^2$ & $0$ & $+1$ \\
$\omega^2$ & $\omega$ &$+1$ & $0$
\end{tabular}\right), 
\end{equation}
 with $\omega = e^{2\pi i/3}$, and $\zeta_{ij} = -\gamma_{ij}^*$. 
In Eq.~\eqref{eq:gamma_eq}, we have expressed 
the $\gamma$ matrix in the $4\times 4$ ``sublattice representation''.
Note: here and in the following appendices, the labels $i$ and $j$ index the four sublattices of a tetrahedron and both $i$ and $j$ run from $1$ to $4$.

\section{EQQ interaction in terms of \texorpdfstring{$O_2$}{TEXT}}
\label{sec:Meqq}

The angular momentum ${\mathbf J}_i$ and Stevens operators $O_2^{\mu}(i)$ in Eqs.~\eqref{eq:HbiEx} and \eqref{eq:Hq} are defined in the \textit{local} frame, while 
${\bm Q}_i$ in Eq.~\eqref{eq:HEQQ} is defined in terms of the components of ${\mathbf J}_i$ in the \textit{global} frame.
It is convenient, however, to express the quadrupole moment ${\bm Q}_i$ in terms of $O_2^\mu$ Stevens operators ($\mu=2,1,0,-1,-2$)
expressed in the \textit{local} frame:
\begin{equation}
\label{eq:Stevens_op}
\begin{split}
O_2^2 &\equiv ({\rm J}^x)^2-({\rm J}^y)^2 , \\
O_2^1 & \equiv \frac{1}{2}({\rm J}^x{\rm J}^z+{\rm J}^z{\rm J}^x) , \\
O_2^{0} &\equiv 3({\rm J}^z)^2-{\rm J}({\rm J}+1),\\
O_2^{-1} &\equiv \frac{1}{2}({\rm J}^y{\rm J}^z+{\rm J}^z{\rm J}^y) , \\
O_2^{-2} & \equiv {\rm J}^x{\rm J}^y+{\rm J}^y{\rm J}^x 
\end{split}
\end{equation}
for each sublattice $i$. 

With these Stevens operators in hand, we express the EQQ Hamiltonian of Eq.~\eqref{eq:HEQQ}
as Eq.~\eqref{eq:Hq}, where the explicit form of the interaction matrix ${\cal M}^{\mu,\nu}_{ij}$ between sublattices $i$ and $j$ ($i,j=1,2,3,4$) is
\begin{equation}
{\cal M}=\frac{1}{32}{\mathscr Q}\cdot\begin{pmatrix*}
\bm 0& \cal A&\cal B&\cal C\\
\cal A&\bm 0&\cal C &\cal B\\
\cal B&\cal C&\bm 0&\cal A\\
\cal C&\cal B&\cal A&\bm 0\\
\end{pmatrix*}.
\end{equation}
Here, $\cal A$, $\cal B$, and $\cal C$ are symmetric $5\times 5$ matrices with the $\mu, \nu$ indices of those matrices 
running in the order of $(2,1,0,-1,-2)$.
\begin{equation*}
{\cal A}=4\left(
\begin{array}{ccccc}
	25 & 28 \sqrt{2} & -3 & 0 & 0 \\
	28 \sqrt{2} & 224 & 12 \sqrt{2} & 0 & 0 \\
	-3 & 12 \sqrt{2} & 17 & 0 & 0 \\
	0 & 0 & 0 & -144 & -24 \sqrt{2} \\
	0 & 0 & 0 & -24 \sqrt{2} & -24 \\
\end{array}
\right),
\end{equation*}
\begin{equation*}
{\cal B}=
\left(
\begin{array}{ccccc}
-47 & 100 \sqrt{2} & 6 & -4 \sqrt{6} & 49 \sqrt{3} \\
100 \sqrt{2} & -208 & -24 \sqrt{2} & -368 \sqrt{3} & 4 \sqrt{6} \\
6 & -24 \sqrt{2} & 68 & 24 \sqrt{6} & 6 \sqrt{3} \\
-4 \sqrt{6} & -368 \sqrt{3} & 24 \sqrt{6} & 528 & -108 \sqrt{2} \\
49 \sqrt{3} & 4 \sqrt{6} & 6 \sqrt{3} & -108 \sqrt{2} & 51 \\
\end{array}
\right),
\end{equation*}
and
\begin{equation*}
{\cal C}=
\left(
\begin{array}{ccccc}
-47 & 100 \sqrt{2} & 6 & 4 \sqrt{6} & -49 \sqrt{3} \\
100 \sqrt{2} & -208 & -24 \sqrt{2} & 368 \sqrt{3} & -4 \sqrt{6} \\
6 & -24 \sqrt{2} & 68 & -24 \sqrt{6} & -6 \sqrt{3} \\
4 \sqrt{6} & 368 \sqrt{3} & -24 \sqrt{6} & 528 & -108 \sqrt{2} \\
-49 \sqrt{3} & -4 \sqrt{6} & -6 \sqrt{3} & -108 \sqrt{2} & 51 \\
\end{array}
\right).
\end{equation*}

\section{Matrix representations of operators in the subspace of two-doublets}
\label{sec:matrix-rep}

From the determination of the crystal field states in Sec.~\ref{subsec:cf-analysis}, the wave functions of the two lowest doublets of \tgo~ are given in Table~\ref{CEFcomp},
\begin{equation}
\begin{gathered}
\vert {\rm 1st}/{\rm 2nd} \rangle
=a_1|\mp 5\rangle\pm b_1| \pm 4\rangle \pm c_1|\mp 2\rangle+d_1|\pm 1\rangle,\\
\vert {\rm 3rd}/{\rm 4th} \rangle
=a_2|\mp 4\rangle\pm b_2|\pm 5\rangle \pm c_2|\mp 1\rangle+d_2|\pm 2\rangle.\\
\end{gathered}
\label{eq:wave}
\end{equation} 
with $(a_1, b_1,c_1,d_1)=(0.5024,0.8271,0.2300,-0.1028)$ and $(a_2, b_2,c_2,d_2)=(0.5296,0.8181,0.1451,-0.1933)$.
The crystal field Hamiltonian in the matrix form for the two-doublet set of states is diagonal and written as,
\begin{equation}
{\cal H}_{\rm cf}={\rm diag}(0,0,\Delta,\Delta),
\label{eq:cf}
\end{equation}
where the first excited doublet lies at an energy $\Delta$ above the ground state~(refer to Table~\ref{CEFcomp}), and here $\Delta=1.5$ meV~(see Table~\ref{CEFenergy}).
Then, from the general form of the states in Eq.~\eqref{eq:wave}, the matrix elements of the various operators of interest can be calculated. For the angular momentum operator within the two doublet Hilbert space (at each site), we have
\begin{equation}
\begin{split}
{\cal P}{\rm J}^x{\cal P}&=t\begin{pmatrix}
0&0&0&1\\
0&0&1&0\\
0&1&0&0\\
1&0&0&0\\
\end{pmatrix},\
{\cal P}{\rm J}^y{\cal P}=t\begin{pmatrix}
0&0&0&i\\
0&0&-i&0\\
0&i&0&0\\
-i&0&0&0\\
\end{pmatrix},\\
{\cal P}{\rm J}^z{\cal P}&=\begin{pmatrix}
j_1&0&j_3&0\\
0&-j_1&0&j_3\\
j_3&0&j_2&0\\
0&j_3&0&-j_2\\
\end{pmatrix},
\end{split}
\label{eq:jj}
\end{equation}
with the projection operator ${\cal P}$ into the reduced doublet-doublet Hilbert space, ${\cal P}\equiv\rm |1st\rangle\langle 1st|+|2nd\rangle\langle 2nd|+|3rd\rangle\langle 3rd|+|4th\rangle\langle 4th|$, for each lattice site and with matrix elements
\begin{equation*}
\begin{split}
j_1&=-5a_1^2+4b_1^2-2c_1^2+d_1^2,\\
j_2&=4 a_2^2-5 b_2^2+c_2^2-2d_2^2,\\
j_3&=5a_1b_2+4a_2b_1-2c_1d_2-c_2d_1,\\
t&=\sqrt{{11}/{2}}(a_1a_2+b_1b_2)+\sqrt{10}(c_1c_2+d_1d_2),
\end{split} 
\end{equation*}
with numerical values 
$(j_1,j_2,j_3,t)=(1.3791,-2.2374,3.8985,2.3696)$.
With the definition of the Stevens operators in Eq.~\eqref{eq:Stevens_op}, we rewrite each of these within the two doublet Hilbert space
\begin{equation}
\begin{split}
{\cal P} O_2^2{\cal P}
&=\begin{pmatrix}
0&A&0&B\\
A&0&-B&0\\
0&-B&0&C\\
B&0&C&0\\
\end{pmatrix},\\
{\cal P}O_2^1{\cal P}
&=\begin{pmatrix}
0&A'&0&B'\\
A'&0&-B'&0\\
0&-B'&0&C'\\
B'&0&C'&0\\
\end{pmatrix},
\end{split}
\label{eq:O22}
\end{equation}
and
\begin{equation}
\begin{split}
{\cal P}O_2^{-2}{\cal P}
&=i\begin{pmatrix}
0&-A&0&-B\\
A&0&-B&0\\
0&B&0&-C\\
B&0&C&0\\
\end{pmatrix},\\
{\cal P}O_2^{-1}{\cal P}
&=i\begin{pmatrix}
0&A'&0&B'\\
-A'&0&B'&0\\
0&-B'&0&C'\\
-B'&0&-C'&0\\
\end{pmatrix},
\end{split}
\label{eq:O2n2}
\end{equation}
and,
\begin{equation}
{\cal P}O_2^{0}{\cal P}
=\begin{pmatrix}
A''&0&B''&0\\
0&A''&0&-B''\\
B''&0&C''&0\\
0&-B''&0&C''\\
\end{pmatrix},
\label{eq:O20}
\end{equation}
with 
\begin{equation*}
\begin{split}
A&=-6\sqrt{30}b_1c_1+21d_1^2,\\
B&=3(\sqrt{30}a_2c_1+7c_2d_1+\sqrt{30}b_1d_2),\\
C&=-21c_2^2+6\sqrt{30}a_2d_2;\\
A'&=\frac{3}{2}(3\sqrt{22}a_1b_1-2\sqrt{10}c_1d_1),\\
B'&=\frac{3}{4}\left(3\sqrt{22}(b_1b_2-a_1a_2)+2\sqrt{10}(d_1d_2-c_1c_2)\right),\\
C'&=\frac{3}{2}(\sqrt{22}a_2b_2-2\sqrt{10}c_2d_2);\\
A''&=33a_1^2+6b_1^2-30c_1^2-39d_1^2,\\
B''&=6a_2b_1-33a_1b_2+39c_2d_1-30c_1d_2,\\
C''&=6d_2^2+33b_2^2-39c_2^2-30d_2^2.
\end{split}
\end{equation*}
Explicitly, we have $(A,B,C)=(-6.0300,-0.9387,-3.8063)$, $(A',B',C')=(8.9952,4.2252,9.3549)$, and $(A'',B'',C'')=(10.4348,-10.1003,21.5581)$. 
We can now plug-in Eqs.~\eqref{eq:cf}--\eqref{eq:O20} 
in Eqs.~\eqref{eq:HbiEx} and \eqref{eq:Hq}, and obtain the matrix representations of the model Hamiltonian ~\eqref{eq:Htot}.

\section{Order parameters}
\label{sec:order-params}

Within a tetrahedron, we define the order parameters for (magnetic) dipolar and (electric) quadrupolar phases as given in Table.~\ref{tab:orders}. For the dipolar phases, we adopt order parameters similar to Ref.~\cite{rau2019frustrated} with the only difference being to replace the pseudospin-$\sfrac{1}{2}$ with the 
mean-field expectation value (henceforth denoted $\langle \ldots \rangle$) of the physical angular moment $\langle {\mathbf J}_i\rangle$. As for the quadrupolar order parameters (see details in Appendix~\ref{sec:matrix-rep}), we first note that
 since $O_2^0=3({\rm J}^z)^2-{\rm J}({\rm J}+1)$ has diagonal matrix elements (see Eq.~\eqref{eq:O20}), $\langle O_2^0\rangle$ is nonzero even in the disordered (paramagnetic) state. Secondly, we observe that, within the pair ($\langle O_2^1\rangle$, $\langle O_2^2\rangle$) or the pair ($\langle O_2^{-1}\rangle$, $\langle O_2^{-2}\rangle$), the expectation value of the two operators within each pair becomes nonzero simultaneously below a critical temperature.
 This can be understood by considering Eqs.~\eqref{eq:O22} and \eqref{eq:O2n2} where we note that $O_2^2$ and $O_2^1$ share the same symmetry while $O_2^{-2}$ and $O_2^{-1}$ have the same symmetry. At the same time,
 the various entries for $O_2^2$ and $O_2^{-2}$ have the same \textit{magnitude} while those for
 $O_2^1$ and $O_2^{-1}$ have the same magnitude. 
 Therefore, a suitable combination of $\langle O_2^1\rangle$ and $\langle O_2^{-1}\rangle$ can be chosen as an order parameter~\cite{mcewen2003anew,lee2018landau} for the quadrupolar ordered phases
of model (\ref{eq:Htot}). As such, we choose the complex scalar $\tau^+=\langle O_2^{1}\rangle+ i\langle O_2^{-1}\rangle$ whose nonvanishing real or imaginary part signals the developing order of one of the two sets of quadrupole operators, ($O_2^1$, $O_2^2$) or ($O_2^{-1}$, $O_2^{-2}$).

	\begin{table*}[t]
	\caption{
			Order parameters for $\mathbf{k}={\mathbf {0}}$ (magnetic) dipole and (electric) quadrupole phases on the pyrochlore lattice. The order parameters are defined in terms of the on-site ($j$=1,2,3,4) ordered magnetic dipole (MD) $\langle {\bf J}_j\rangle$ and electric quadrupole (EQ) $\langle O_2^{\pm 1}(j)\rangle$ moments, respectively.}
		\begin{tabular}{c c|c c}
		\hline\hline	\multicolumn{2}{c}{Dipolar order parameter}&\multicolumn{2}{c}{Quadrupolar order parameter}\\
			\midrule
			Names & Definition in terms of local magnetic moments~\cite{rau2019frustrated} & Names & \begin{tabular}{c} Definition ito. local \\quadrupole moment\end{tabular} \\
			\hline
			\begin{tabular}{c}
				$\Gamma_3$\\ All-in/All-out\\
			\end{tabular}
			& 	$m_{\Gamma_{3}}\equiv$ 	\footnotesize $ (\sigma^z_1+\sigma^z_2+\sigma^z_3+\sigma^z_4)$ & 	\begin{tabular}{c}
				Ferro-\\quadrupole \\
			\end{tabular}&	$q_{\rm FQ}\equiv$ 	$(\tau_1^++\tau_2^++\tau_3^++\tau_4^+)$\\ \hline
			\begin{tabular}{c}
				$\Gamma_{9-\rm SI}$ \\
				Ordered\\spin ice 
			\end{tabular} &$\bm m_{\Gamma_{9-\rm SI}}\equiv$ \footnotesize $ \left(\begin{array}{c}
			\sigma^z_1+\sigma^z_2-\sigma^z_3-\sigma^z_4\\
			\sigma^z_1-\sigma^z_2+\sigma^z_3-\sigma^z_4\\
			\sigma^z_1-\sigma^z_2-\sigma^z_3+\sigma^z_4 
			\end{array}\right)$ 
			& \begin{tabular}{c}
				Antiferro- \\ quadrupole \\
			\end{tabular}&$\bm q_{\rm AFQ}\equiv$ $\begin{pmatrix}
			\tau_1^++\tau_2^+-\tau_3^+-\tau_4^+\\
			\tau_1^+-\tau_2^++\tau_3^+-\tau_4^+\\
			\tau_1^+-\tau_2^+-\tau_3^++\tau_4^+
			\end{pmatrix}$\\ \hline
			\cline{3-4}
			\begin{tabular}{c}
				$\Gamma_5$ \\
				Antiferro- \\magnet\end{tabular}
			&$\bm m_{\Gamma_5}\equiv$ \footnotesize $\left(\begin{array}{c}
			\sigma^x_1+\sigma^x_2+\sigma^x_3+\sigma^x_4\\ 
			\sigma^y_1+\sigma^y_2+\sigma^y_3+\sigma^y_4
			\end{array}\right) $
			& \multirow{3}{5em}{\\$\quad$\\$\quad$On-site \\$\quad$ordered\\ $\;\;\,$ moments}&\multirow{3}{10em}{\begin{tabular}{c}\\$\,$\\$\bm\sigma_j\equiv \langle {\bf J}_j\rangle$\vspace{+01mm}\\$\tau^+_j \equiv \langle O_2^1 (j)\rangle+i\langle O_2^{-1} (j)\rangle$\vspace{+01mm}\\$j=1,2,3,4$\vspace{+01mm}\\
					\end{tabular}}\\\cline{1-2}
			\begin{tabular}{c}
				$ \Gamma_7$\\Palmer-\\Chalker\\
			\end{tabular} &	$\bm m_{\Gamma_7}\equiv$ \footnotesize $ \left(\begin{array}{c}
			\sigma^y_1 + \sigma^y_2 -\sigma^y_3 -\sigma^y_4\\
			\left(-\frac{\sqrt{3}}{2} \sigma^x_1-\frac{1}{2}\sigma^y_1\right)-
			\left(-\frac{\sqrt{3}}{2} \sigma^x_2-\frac{1}{2}\sigma^y_2\right)+
			\left(-\frac{\sqrt{3}}{2} \sigma^x_3-\frac{1}{2}\sigma^y_3\right)-
			\left(-\frac{\sqrt{3}}{2} \sigma^x_4-\frac{1}{2}\sigma^y_4\right)
			\\
			\left(\frac{\sqrt{3}}{2} \sigma^x_1-\frac{1}{2}\sigma^y_1\right)-
			\left(\frac{\sqrt{3}}{2} \sigma^x_2-\frac{1}{2}\sigma^y_2\right)-
			\left(\frac{\sqrt{3}}{2} \sigma^x_3-\frac{1}{2}\sigma^y_3\right)+
			\left(\frac{\sqrt{3}}{2} \sigma^x_4-\frac{1}{2}\sigma^y_4\right) 
			\end{array}\right)$
			&\\ \cline{1-2}
			\begin{tabular}{c}
				$ \Gamma_{9-\rm SFM}$\\
				Splayed-\\ferromagnet\\
			\end{tabular} & $\bm m_{\Gamma_{9-\rm SFM}}\equiv$ \footnotesize $\left(
			\begin{array}{c}
			\sigma^x_1 + \sigma^x_2 -\sigma^x_3 -\sigma^x_4\\
			\left(-\frac{1}{2}\sigma^x_1+\frac{\sqrt{3}}{2} \sigma^y_1\right)-
			\left(-\frac{1}{2}\sigma^x_2+\frac{\sqrt{3}}{2} \sigma^y_2\right)+
			\left(-\frac{1}{2}\sigma^x_3+\frac{\sqrt{3}}{2} \sigma^y_3\right)-
			\left(-\frac{1}{2}\sigma^x_4+\frac{\sqrt{3}}{2} \sigma^y_4\right)\\
			\left(-\frac{1}{2}\sigma^x_1-\frac{\sqrt{3}}{2} \sigma^y_1\right)-
			\left(-\frac{1}{2}\sigma^x_2-\frac{\sqrt{3}}{2} \sigma^y_2\right)-
			\left(-\frac{1}{2}\sigma^x_3-\frac{\sqrt{3}}{2} \sigma^y_3\right)+
			\left(-\frac{1}{2}\sigma^x_4-\frac{\sqrt{3}}{2} \sigma^y_4\right) 
			\\
			\end{array}
			\right)
			$ 
			\\ 
			\hline\hline
		\end{tabular}
		\label{tab:orders} 
			\end{table*}

\section{MF-RPA approximation}
\label{sec:MF-RPA}

\subsection{Mean-field approximation}

We use mean-field theory to determine the equilibrium phases of the model Hamiltonian given by Eq.~\eqref{eq:Htot}. The mean-field (MF) Hamiltonian acting at sublattice $i$ is
\begin{equation}
{\cal H}_{\rm MF}(i)=2\sum_\alpha h_{\rm MD}^\alpha(i) \cdot {\rm J}_i^\alpha+2\sum_\mu h_{\rm EQ}^\mu(i)\cdot O_2^\mu(i),
\label{eq:hmf}
\end{equation}
with the molecular field induced by the ordering of magnetic dipolar (MD) and electric quadrupolar (EQ) moments, 
\begin{equation}
h_{\rm MD}^\alpha(i)=\sum_{j,\beta}{\cal J}_{ij}^{\alpha\beta}\langle {\rm J}_j^\beta\rangle,\quad
h_{\rm EQ}^\mu(i)=\lambda\sum_{j,\nu}{\cal M}_{ij}^{\mu\nu}\langle O_2^\nu(j)\rangle.
\label{eq:hfield}
\end{equation}
Here, we are foremost interested in $\mathbf{k}={\mathbf {0}}$ phases, where, again, labels $i,j=1,2,3,4$ index the four sublattices. There is a prefactor of 2 introduced in front of the two summations in Eq.~\eqref{eq:hmf} because the six nearest neighbors 
on the pyrochlore lattice amount to three pairs of equivalent sites 
in a $\mathbf{k}={\mathbf {0}}$ phase. Here, $\alpha,\beta=x,y,z$ labels the three components of the MD moment and $\mu,\nu=0, \pm 1, \pm 2$ labels the five components of EQ moment, as described by
the $O_2^\mu$ Stevens operators (see Appendices \ref{sec:Meqq}
and \ref{sec:matrix-rep}). The free energy per site, $f$, is given by
\begin{equation}
4f=-k_{\rm B}T\ln Z_{\rm mf} -\sum_{(i,j)}\left ( {\cal J}_{ij}^{\alpha\beta}\langle J_i^\alpha\rangle\langle J_j^\beta\rangle+\lambda {\cal M}_{ij}^{\mu\nu}\langle O_2^\mu(i )\rangle\langle O_2^\nu(j) \rangle \right ) ,
\label{eq:free-energy}
\end{equation}
where $Z_{\rm mf}$ is the ``partition function'' associated with Eq.~\eqref{eq:hmf}, defined below in Eq.~\eqref{eq:Zmf}, and $k_{\rm B}$ is the Boltzmann constant. We solve Eqs.~\eqref{eq:hmf} and ~\eqref{eq:hfield} in a self-consistent manner: starting from a random configuration of $\langle {\rm J}_i^\alpha\rangle$, $\langle O_2^\mu(i)\rangle$, the molecular field \eqref{eq:hfield} and single-site mean-field Hamiltonian ~\eqref{eq:hmf} can be calculated. Diagonalizing this Hamiltonian yields the eigen-energies $E_{i,a}$ and eigenstates $|\phi_{i,a}\rangle$ ($i$ labels each of the four sublattices and $a$ labels the four eigenstates within the 4-state 
 Hilbert subspace, $a=1, \ldots 4$). We then compute $Z_{\rm mf}$ and the expectation value of operators $P_i$, with $P_i \equiv ({\rm J}_i^\alpha, O_2^\mu(i))$ (giving eight operators in total for each sublattice $i$; $\alpha=x,y,z$ and $\mu = 2, 1, 0, -1, -2$.
 \begin{equation}
 \label{eq:Zmf}
 \begin{split}
 Z_{\rm mf}&=\sum_{i=1}^{4} \, \sum_{a=1}^{4} e^{-\beta E_{i,a}},\\
\langle P_i\rangle&=\sum_{a=1}^{4} \frac{e^{-\beta E_{i,a}}}{Z_{\rm mf}}\langle \phi_{i,a}|P_i| \phi_{i,a}\rangle.
 \end{split}
 \end{equation}
 Performing this calculation for all four sublattices, we determine the various 
 $\langle P_i\rangle$ which are then used as the input for a new MF Hamiltonian, with this process being repeated until convergence is achieved. The number of iterations necessary to reach convergence depends on temperature and location in the Hamiltonian~\eqref{eq:Htot} parameter space. Usually, a larger number of iterations is required when the system is near a phase boundary. After convergence has been achieved, the free energy \eqref{eq:free-energy} is calculated. For each set of parameters, the self-consistent processes is repeated for hundreds of times, and we select the solution that minimizes the free energy. Eventually, we obtain phase diagrams at both ``zero'' ($T=10^{-7}$ K) and finite temperatures, as shown in Fig.~\ref{fig:pd+ins}(a,b).

\subsection{Spin dynamics \& random phase approximation}
\label{app:spin_dynamics}

Having determined the equilibrium mean-field (MF) phases, we can then compute the associated dynamical susceptibility using the random phase approximation (RPA). Upon convergence of the mean-field iterations and determination of the minimal free-energy $F$ in Eq.~\eqref{eq:free-energy}, we compute the single-site MF magnetic susceptibility $\bm\chi_i^0(\omega)$,
 \begin{equation}
 \chi_{i}^{0,\alpha\beta}(\omega)=\sum_{a,b}
 \frac{{M}_{ab}^{\alpha}(i){ M}_{ba}^{\beta}(i)}{E_{i,a}-E_{i,b}-\omega}(n_{i,a}-n_{i,b}).
 \label{eq:sus}
 \end{equation}
 Here, ${M}_{ab}^{\alpha}(i) \equiv \langle \phi_{i,a}| J^\alpha_i-\langle J^\alpha_i\rangle |\phi_{i,b} \rangle $ is the matrix element of $J_i^\alpha$ between two mean-field states $|\phi_{i,a}\rangle$ and $|\phi_{i,b}\rangle$, with associated eigen energy $E_{i,a}$ and $E_{i,b}$, respectively, and
 $n_{i,a}=e^{-\beta E_{i,a}}/Z_{\rm mf}$.
 To access the dynamic (magnetic) susceptibility, $\bm\chi(\bm q,\omega)$, in the absence of quadrupole-quadrupole interactions ($\lambda=0$), one has the site-dependent (pairwise) RPA equation~\cite{jensen1991},
 \begin{equation}
\chi_{ij}^{\alpha\beta}(\bm q,\omega)=\delta_{ij}\chi_{i}^{0,\alpha\beta}(\omega)+\sum_{\gamma,\delta,k}\chi_{i}^{0,\alpha\gamma}(\omega){\cal J}_{ik}^{\gamma\delta}(\bm q)\chi_{kj}^{\delta\beta}(\bm q,\omega).
\label{rpa}
 \end{equation}
 Here, $\alpha,\beta, \gamma,\delta=x,y,z$ and $i,j,k=1,2,3,4$, and ${\cal J}_{ik}^{\gamma\delta}(\bm q)$ is the Fourier transform of the bilinear exchange. When there are solely bilinear interactions between the $\mathbf J$ moments, Eqs.~\eqref{eq:sus} and \eqref{rpa} are the only ones needed.
 
 For a model also taking account of the interactions between multipoles, such as the EQQ interaction in Eq.~\eqref{eq:Htot}, we need to compute a generalized multipolar susceptibility~\cite{rau2016order}, $\bm X(\bm q,\omega)$. Similarly to Eq.~\eqref{eq:sus}, the single-site multipolar susceptibility is now
 \begin{equation}
{X}_{i}^{0,\alpha\beta}(\omega)=\sum_{a,b}
 \frac{{\bm M}_{ab}^{\alpha}(i){\bm M}_{ba}^{\beta}(i)}{E_{i,a}-E_{i,b}-\omega}(n_{i,a}-n_{i,b}) .
 \label{eq:sus-ge}
\end{equation}
Here, ${\bm M}_{ab}^{\alpha}(i) \equiv \langle \phi_{i,a}| P_i^\alpha-\langle P^\alpha_i\rangle |\phi_{i,b} \rangle $, with ``expanded'' labels $\alpha=1,2, \ldots 8$, corresponding to the eight components of the generalized operator $ {\bm P}_i=\left(J_i^x, J_i^y, J_i^z; O_2^{2}(i),O_2^{1}(i),O_2^{0}(i),O_2^{-1}(i),O_2^{-2}(i)\right)$. The generalized multipolar susceptibility $\bm X(\bm q,\omega)$ can then be determined from the RPA equations,
\begin{equation}
X_{ij}^{\alpha\beta}(\bm q,\omega)=\delta_{ij}X_{i}^{0,\alpha\beta}(\omega)+\sum_{\gamma,\delta,k}X_{i}^{0,\alpha\gamma}(\omega){\cal G}_{ik}^{\gamma\delta}(\bm q)X_{kj}^{\delta\beta}(\bm q,\omega),
\label{rpa-ge}
\end{equation}
where ${\cal G}_{ij}(\bm q)$ is the Fourier transform of the multipolar interaction matrix ${\cal G}_{ij}$, 
\begin{equation}
{\cal G}_{ij}=\begin{pmatrix}
{\cal J}_{ij}&\bm 0\\
\bm 0&{\cal M}_{ij}\\
\end{pmatrix}.
\end{equation}

To solve Eq.~\eqref{rpa-ge} numerically, we carry the following replacement, $ \omega\to \omega+i\epsilon$, with the non-zero value (positive) $\epsilon$ made to correspond to the instrumental energy resolution~\cite{jensen1991}, choosing here $\epsilon=0.02$ meV. The multipolar susceptibility $\bm X(\bm q,\omega)$ is an $8\times 8$ matrix, and includes the ($3\times 3$ sub-block) magnetic susceptibility $\bm \chi (\bm q,\omega)$, whose imaginary part is directly related to the intensity measured in an inelastic neutron scattering experiment~\cite{jensen1991}.

\subsection{Spin dynamics \& neutron scattering structure factor}
 
In our manuscript, we defined the unpolarized partial differential cross section for the neutron scattering intensity as
 \begin{equation}
 \begin{split}
I(\bm Q,\omega)=&I_0\frac{|\bm k'|}{|\bm k|}\frac{|F(\bm Q)|^2}{1-e^{-\beta\omega}}\sum_{\alpha\beta}(\delta_{\alpha\beta}-\hat{Q}_\alpha\hat{Q}_\beta)\times\\&\sum_{i,j; \gamma,\delta} U_{\alpha,\gamma}^{i} U_{\beta,\delta}^{j} {\rm Im}\;\left[e^{-i\bm G\cdot(\bm r_i-\bm r_j)}\chi_{ij}^{\gamma\delta}(\bm q,\omega)\right],
 \end{split}
 \end{equation}
where, again, $\chi_{ij}^{\gamma\delta}$ are the elements of the magnetic susceptibility $\chi(\bm q,\omega)$ where $\bm Q=\bm q+\bm G$, $\bm G$ is a reciprocal lattice vector of the FCC lattice, and $\bm q$ is a wave vector inside the first Brillouin zone. $F(\bm Q)$ is the magnetic form factor of Tb$^{3+}$~\cite{Browntable}. $\bm r_i$ denotes the position of sublattice site $i$. $U_{\alpha,\gamma}^i$ is the rotation matrix from the local ($\gamma$) frame to the global frame ($\alpha$) and $I_0$ is an arbitrary overall scale factor. $\bm k'$ and $\bm k$ are the initial and final neutron momenta and ${|\bm k'|}/{|\bm k|}=(1-\omega/E_{\rm i})$~\cite{rau2016anisotropic}, which reduces the relative intensity of higher energy lying features, and where $E_{\rm i}=3.3$ meV is the neutron incident energy. The powder averaged cross section is given by
\begin{equation}
I_{\rm avg}(Q,\omega)=\int d\hat{\bm Q} \,I(Q\hat{\bm Q}, \omega).
\end{equation}
As shown in Fig.~\ref{fig:pd+ins}, at low temperatures, $I_{\rm avg}(Q,\omega)$ exhibits
dispersive features in both the crystal field transition and the low-energy mode centered 1.5 and 0.18 meV, respectively, as a result of the anisotropic bilinear exchange and EQQ interactions.

\section{Choice of Hamiltonian parameter values}
\label{sec:fit}

\subsection{\texorpdfstring{{${\cal J}_{zz}$}}{TEXT} bilinear coupling parameter}
\label{sec:Jzz_coupling}

Notwithstanding that the toy-model of Eq.~\eqref{eq:Htot} is already an important simplification of the most general multipolar Hamiltonian~\cite{rau2019frustrated} one would expect for Tb$_{2}$Ge$_{2}$O$_{7}$, even a somewhat accurate determination of ``only'' the five parameters $\{{\cal J}_{zz}, {\cal J}_{\pm}, {\cal J}_{z\pm}, 
{\cal J}_{\pm\pm},\lambda \}$ by fitting the experimental data would be a challenging task at the present time. To appreciate this, one may consider the significant theoretical and experimental efforts spanning nearly ten years that have been required to determine the equivalent of those couplings (whose projection in the crystal field ground doublet yield an effective pseudospin-$1/2$ model) for Yb$_2$Ti$_2$O$_7$~\cite{thompson2017quasiparticle,scheie2019multiphase}, Yb$_2$Ge$_2$O$_7$~\cite{sarkis2019Yb2Ge2O7} and Er$_2$Ti$_2$O$_7$~\cite{savary2012order}, and where the added complexities of low-lying excited crystal field levels (as is the case for terbium pyrochlores) does not exist~\footnote{That may not be exactly true for Er$_2$Ti$_2$O$_7$. See Ref.~\cite{rau2016order}.}. In addition, such a procedure was only possible for those three compounds thanks to the availability of large high-quality single crystals. As discussed in the main manuscript, Tb$_2$Ge$_2$O$_7$ is currently available through high pressure synthesis only, and exclusively as polycrystalline samples.

From past experience with Yb$_2$Ti$_2$O$_7$, Yb$_2$Ge$_2$O$_7$ and Er$_2$Ti$_2$O$_7$, as well as the strongly Ising~\cite{rau2015quantumeffects} Dy$_2$Ti$_2$O$_7$ spin ice material~\cite{yavorskii2008emergent,henelius2017refrustration}, one has found the Ising pseudospin-$1/2$ coupling $J_{zz}$ in model~\eqref{H_eff} typically of the order of $0.1$ meV. 
Writing 
\begin{equation*}
J_{zz} \sim {\cal J}_{zz}
\vert \langle {\rm 1st} \vert {\rm J}^z \vert {\rm 1st} \rangle \vert^2 ,
\end{equation*}
and using the spectral decomposition of $\vert {\rm 1st} \rangle$ from Table~\ref{CEFcomp}, we get
$J_{zz} \sim 1.91 {\cal J}_{zz}$, giving ${\cal J}_{zz} \sim 0.05$ meV.

The mean-field theory approximation that we employ to compute the critical temperatures for quadrupolar and dipolar ordering (see Fig.~\ref{fig:pd+ins}(b)) neglects thermal and quantum fluctuations and over-estimates the critical temperatures. 
As such, in order to obtain a temperature scale for those orderings that is roughly of the order of the temperature at which the growth rate of the collective paramagnetic correlations peaks in Tb$_2$Ge$_2$O$_7$, as signaled by the heat capacity bump at $T^*\sim 1.1$ K (see Fig.~\ref{Tb_Cp}), we deflate the above ${\cal J}_{zz}$ scale by a factor $4$. This gives ${\cal J}_{zz} \sim 0.012$ meV, the value we consider throughout the paper. As found in Fig.~\ref{fig:pd+ins}(b)), this choice for ${\cal J}_{zz}$ gives a 
``reasonable'' mean-field critical temperature $T_c^{\rm mf}\sim 0.8$ K $\pm 0.1$ K for the various dipolar and quadrupolar orders of Fig.~\ref{fig:pd+ins}(a), along with pinning the overall energy scale of our model needed to get energy scales (gap and bandwidth of $I_{\rm avg}$ (see Figs.~\ref{fig:pd+ins}(c,d)) compatible with the experimental ones (see Fig.~\ref{DCS_Slices}).

\begin{figure}[ht!]
 	\begin{overpic}[width=16pc]{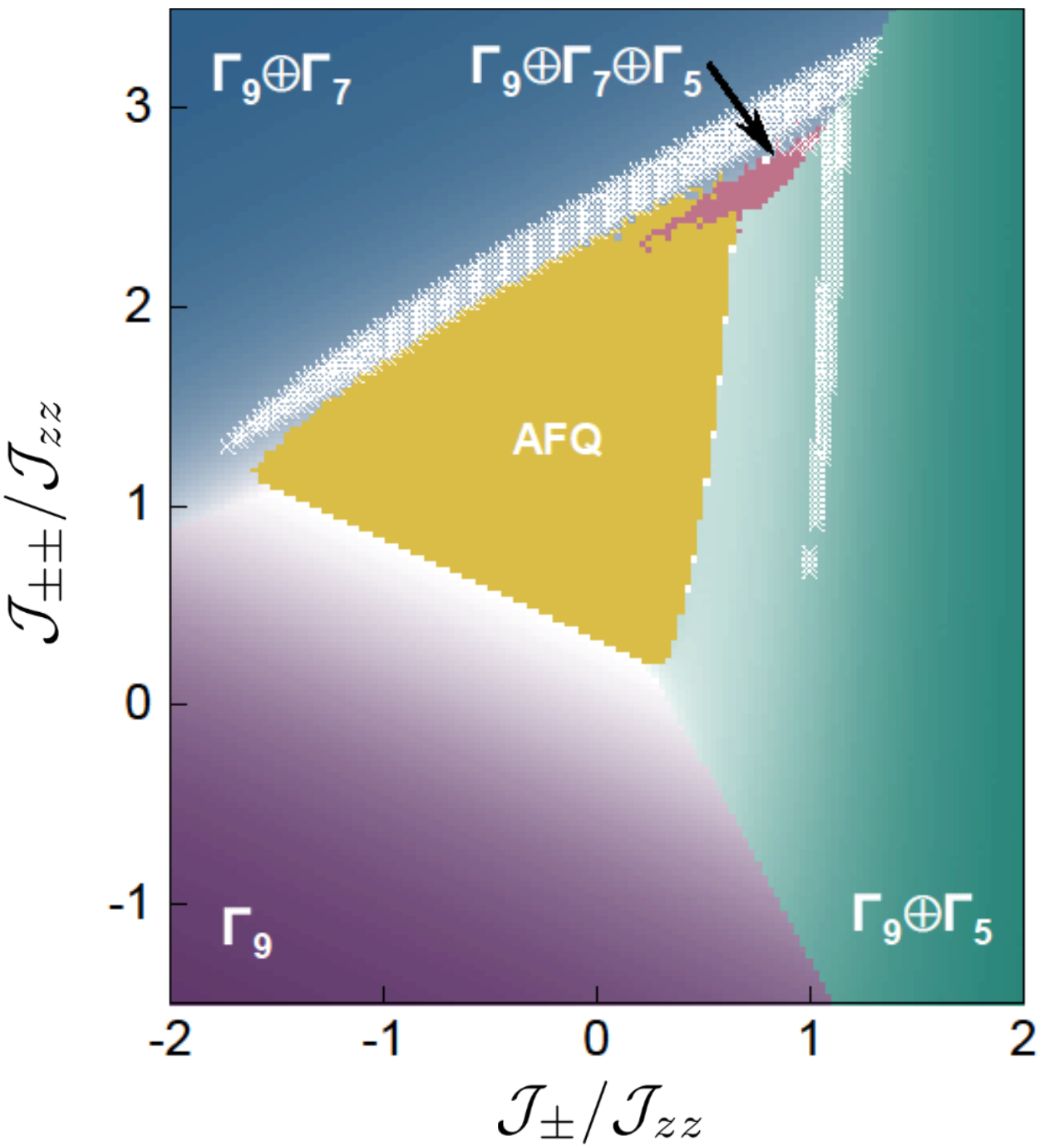}
 	\end{overpic}
 	\caption{Rough fitting of both splitting of ground doublet $E_{\rm gap}=0.18$ meV and canting angle $\alpha=24.5^{\circ}$ with ``tolerances'' taken as $\pm 0.015$ meV and $\pm 1.5^{\circ}$, respectively. Here ${\cal J}_{zz }=0.012$ meV, ${\cal J}_{z\pm}=(2/3){\cal J}_{zz }$, and $\lambda=0.28$. This rough fitting places possible parameters, which satisfy the above two constraints simultaneously, by white crosses in the $\Gamma_9\oplus\Gamma_7$ and $\Gamma_9\oplus\Gamma_5$ regions.}
 	\label{fig:fit}
 \end{figure}

\subsection{
\texorpdfstring{{${\cal J}_{uv} \ne {\cal J}_{zz}$}}{TEXT} 
bilinear and 
\texorpdfstring{{${\lambda}$}}{TEXT} 
quadrupolar coupling parameters}
\label{sec:Juv_couplings}

We aim to produce a two-dimensional ground state phase diagram with two dimensionless bilinear couplings (${\cal J}_{uv}'/{\cal J}_{zz}$ and ${\cal J}_{uv}''/{\cal J}_{zz}$), 
and leave one ${\cal J}_{uv}'''$ to be determined along with the quadrupolar scale factor $\lambda$. We choose ${\cal J}_{uv}'\equiv {\cal J}_{\pm}$ and
${\cal J}_{uv}''\equiv {\cal J}_{\pm\pm}$, and proceed to estimate the third one, ${\cal J}_{uv}'''\equiv {\cal J}_{z\pm}$.

Two quantities describing Tb$_2$Ge$_2$O$_7$ are particularly convenient to consider to
constrain some of the parameters of Eq.~\eqref{eq:Htot}: the energy gap characterizing the splitting of the ground doublet, $E_{\text{gap}}=0.18\pm 0.015$~meV, and the canting angle $\alpha = 24.5 \pm 1.5^{\circ}$ of the Tb magnetic moments away from their local $\langle 111 \rangle$ direction. Using these two values as constraint, we perform a two-parameter fitting (for a range of ${\cal J}_{\pm}/{\cal J}_{zz}$ and ${\cal J}_{\pm\pm}/{\cal J}_{zz}$) that incorporates dipolar ordered phases surrounding a central quadrupolar (AFQ) phase lacking magnetic dipolar order.

Let us expand on the general procedure used. The ground state phase diagram of model \eqref{eq:Htot} in the ${\cal J}_{\pm}/{\cal J}_{zz}$ and ${\cal J}_{\pm\pm}/{\cal J}_{zz}$ space parametrically evolves as the independent ${\cal J}_{z\pm}$ and $\lambda$ quantities are varied. For each such $({\cal J}_{z\pm},\lambda)$ pair, one can trace in the $({\cal J}_{\pm}/{\cal J}_{zz},{\cal J}_{\pm\pm}/{\cal J}_{zz})$ plane two ``contour curves'' that correspond to the constrained values $\alpha$ and $E_{\text{gap}}$ (and of width corresponding to twice the aforementioned respective tolerance $\delta\alpha=1.5^{\circ}$, $\delta E_{\text{gap}} = 0.015$ meV). In essence, our goal is to find region(s) in the $({\cal J}_{\pm}/{\cal J}_{zz},{\cal J}_{\pm\pm}/{\cal J}_{zz})$ phase diagram for a $({\cal J}_{z\pm},\lambda)$ pair where those contours overlap significantly.

 First, we comment on how $\alpha$ and $E_{\text{gap}}$ are determined in the MF-RPA calculations. The canting angle $\alpha$ is set by the orientation of the magnetic moments,$\langle {\mathbf J} \rangle$, which are determined by solving the mean-field equations~\eqref{eq:hmf} and \eqref{eq:hfield}. Thus, $\alpha$ depends on both the magnetic couplings ${\cal J}_{uv}$ and the scale of the quadrupolar interactions, $\lambda$. In the course of the calculations, a converged mean field spectrum, obtained by diagonalizing the mean field Hamiltonian~\eqref{eq:hmf}, ${\cal H}_{\rm mf}(i) |\phi_{i,a}\rangle=E_{i,a}|\phi_{i,a}\rangle$, is ultimately obtained. To simplify and speed up the calculations, foregoing a full calculation of $I_{\rm avg}(Q,\omega)$ in four-dimensional $({\cal J}_{\pm}/{\cal J}_{zz},{\cal J}_{\pm\pm}/{\cal J}_{zz},J_{z,\pm},\lambda)$ parameter space, we take the splitting of the doublet in the (quadrupolar or dipolar) ordered phases at $T=0^+$ to match $E_{\rm gap}=0.18$ meV. This is an approximation but, within the aforementioned tolerance $\delta E_{\text{gap}} = 0.015$ meV, it does capture the region of strongest inelastic intensity at $E_{\text{gap}} \approx 0.18$ meV (see Fig.~\ref{DCS_Slices}(i-l)) that we aim to fit. 
 This approximation is further justified by considering the weakly dispersive $I(Q,\omega)$ where the experimental gap at $Q\rightarrow 0^+$, which is perhaps a more direct signature of the split-doublet (in a mean-field picture), is also $\sim 0.18$ meV, as is the gap at $Q=1.2 {\AA}^{-1}$. More to the point, the splitting of the doublet 
 is induced by the combined dipolar (${\bm h}_{\rm MD}$) and quadrupolar (${\bm h}_{\rm EQ}$) molecular fields in Eq.~\eqref{eq:hfield}, which both depend on the magnitude and direction of the moments, $\langle {\mathbf J} \rangle$, and thus, implicitly, on the canting angle $\alpha$ in the dipolar ordered phases of the model. Conversely, in the AFQ phase 
 found in the model, while the non-Kramers doublet is split by the off-diagonal ${\bm h}_{\rm EQ}$, there is no dipolar order and $\alpha$ is undefined.

 Summarizing the paragraph above, while $\alpha$ and $E_{\rm gap}$ are two distinct physical quantities, they are tied together, albeit in a non-trivial (\emph{e.g.} ``nonlinear'') manner. Therefore, for an arbitrary $({\cal J}_{z\pm},\lambda)$ pair, the aforementioned two contour curves do not generally overlap, but when they do overlap, the overlap region evolves upon varying ${\cal J}_{z\pm}$ and $\lambda$. In the limit of large $\lambda$ or small ${\cal J}_{z\pm}$, there is no dipole orders (for a range of ${\cal J}_{\pm}/{\cal J}_{zz}$ and ${\cal J}_{\pm\pm}/{\cal J}_{zz}$); while in the opposite limit, small $\lambda$ or large ${\cal J}_{z\pm}$, pure quadrupolar order is absent. As a result, what we generally observe is that $\alpha(J_{z\pm},\lambda)$ and $E_{\rm gap}(J_{z\pm},\lambda)$ reach their minimum value on the periphery of the AFQ phase (yellow region in Fig.~\ref{fig:fit}), but at different rate depending on the direction upon approaching the AFQ phase in the 
 $({\cal J}_{\pm}/{\cal J}_{zz}, {\cal J}_{\pm\pm}/{\cal J}_{zz}$) plane for a given $(J_{z\pm},\lambda)$ pair. Ultimately, we find for ${\cal J}_{z\pm} \approx (2/3){\cal J}_{zz}$ and $\lambda \approx 0.28$ two candidate regions with significant overlap for the fitted $\alpha$ and $E_{\rm gap}$ values -- those are indicated by the white wedges in Fig.~\ref{fig:fit}. We therefore set ${\cal J}_{z\pm}=(2/3) {\cal J}_{zz}$ and $\lambda=0.28$.

\subsection{Candidate location of 
\texorpdfstring{Tb$_{2}$Ge$_{2}$O$_{7}$}{TEXT} in phase diagram}
\label{sec:candidate_loc}

 To proceed, one needs to further tighten the choice of the two other bilinear parameters, ${\cal J}_{\pm\pm}$ and ${\cal J}_{\pm}$, within the white wedges in Fig.~\ref{fig:fit} in order to relate our model calculations to the experimental results on \tgo{}. In this work we focus on the ``upper'' (slanted) boundary that runs from 
(${\cal J}_{\pm\pm}/{\cal J}_{zz}\sim 1.2$, ${\cal J}_{\pm}/{\cal J}_{zz} \sim -1.5$) to (${\cal J}_{\pm\pm}/{\cal J}_{zz}\sim 3$, ${\cal J}_{\pm}/{\cal J}_{zz} \sim 1.5$) as opposed to the one that runs almost vertically at $\sim {\cal J}_{\pm}/{\cal J}_{zz}\sim 1.2$ within the 
 $\Gamma_9 \oplus\Gamma_5$ phase (from ${\cal J}_{\pm\pm}/{\cal J}_{zz}\sim 0.5$ to ${\cal J}_{\pm\pm}/{\cal J}_{zz}\sim 2.8$). 
 Our key reason for focusing on this boundary is motivated by the interpretation of the experimental results that strongly suggest that \tgo{} finds itself naturally tuned closed to a phase boundary separating quadrupolar (EQ) and magnetic dipolar (MD) orders. The same argument may also apply to Tb$_2$Ti$_2$O$_7$~\cite{takatsu2016quadrupole}. The vertical white wedge being fairly well removed from the AFQ and the $\Gamma_9 \oplus\Gamma_5$ boundary thus seems to be a {\it less} suitable candidate for this first systematic study of \tgo{}.

Considering now the upper white wedge boundary, riding in the 
$\Gamma_9 \oplus\Gamma_7$ dipole ordered phase, we wish to pick a location in that boundary where we can superpose the $I_{\rm avg}(Q,\omega)$ originating from two nearby competing MD and EQ phases -- \emph{e.g.} the $\eta$-weighting superposition illustrated in Fig.~\ref{fig:pd+ins}(c,d).
In that $\Gamma_9 \oplus\Gamma_7$-AFQ boundary, we select the pair of points \#1 and \#2 in the upper right corner, with point \#1 in the MD ordered $\Gamma_9 \oplus\Gamma_7$ and point \#2 just across the boundary in the narrow yellow sliver of AFQ phase above the $\Gamma_9 \oplus\Gamma_7 \oplus \Gamma_5$ phase -- see Fig.~\ref{fig:pd+ins}(a). The physical reason for considering a point at the upper top/right limit of that boundary is non-trivial and goes as follows. We find in our calculations that the quadrupolar molecular field, ${\bm h}_{\rm EQ}$ (induced by the predominant MD order) gets progressively \emph{weaker} within the upper white wedge as one goes from the top right to the bottom left (near ${\cal J}_{\pm\pm}/{\cal J}_{zz}\sim 1.2$, ${\cal J}_{\pm}/{\cal J}_{zz} \sim -1.5$). We noted in Fig.~\ref{DCS_Slices}(c-f) that the intensity of the mode of energy $E_{\rm gap}\sim 0.18$ meV at $Q\sim 1.2\AA^{-1}$ is significantly more intense than the crystal field exciton at energy $\sim 1.5$ meV. In a MF+RPA description, the intensity of this collective mode at $Q\sim 1.2\AA^{-1}$ and 0.18 meV is controlled by the off-diagonal part of the molecular quadrupolar field ${\bm h}_{\rm EQ}$ splitting the ground crystal field doublet (and thus entangling \ket{\rm 1st} and \ket{\rm 2nd}), recalling that for a non-Kramers ion such as Tb$^{3+}$, the time-odd (\emph{e.g.} MD) molecular field does not mix the 
\ket{\rm 1st} and \ket{\rm 2nd} (to zeroth order in $1/\Delta$), making the inelastic neutron scattering signal of this collective mode essentially invisible. Thus aiming to obtain the strongest INS intensity of the collective mode within the $\Gamma_9 \oplus\Gamma_7$ phase, we consider the farthest possible point in the upper white wedge of Fig.~\ref{fig:fit} -- point \#1. A superposition of the corresponding $I_{\rm avg}(Q,\omega)$ for a point (\emph{i.e.} ${\cal J}_{\pm\pm}/{\cal J}_{zz}$ and ${\cal J}_{\pm}/{\cal J}_{zz}$ parameter values) across the boundary and in the AFQ phase (\emph{e.g.} point \#2), as described in the caption of Fig.~\ref{fig:pd+ins} and Section~\ref{subsec:INS-theory}, produces a theoretical inelastic neutron scattering signal at energy $\lesssim 0.2$ meV similar to the experimental one (compare Fig.~\ref{DCS_Slices} and Fig.~\ref{fig:pd+ins}(c,d) for $\eta\sim 0.6$).
We note, in passing, that this whole rationale leads us to consider a region in the phase diagram of our toy-model which is (very) close to the aforementioned $\Gamma_9 \oplus\Gamma_7 \oplus \Gamma_5$ phase that has unequal $\langle {\mathbf{J}}_i\rangle$ for each of the four sublattices and which, as discussed in Section ~\ref{subsec:mft-phase-diagram}, may be an indicator that a more sophisticated calculation relaxing the imposed ${\mathbf k}={\mathbf 0}$ solution might actually find ${\mathbf k}\ne {\mathbf{0}}$. Further work is necessary to explore this possibility.

\end{document}